\documentclass[twocolumn]{aastex631}

\usepackage{amsmath}	
\usepackage{amsfonts}
\usepackage{bm}
\usepackage[T1]{sansmath} 
\usepackage{graphicx}
\begin{document}

\title{Favorable conditions for heavy element nucleosynthesis in rotating proto-magnetar winds}

\author{Tejas Prasanna}
\affiliation{Department of Physics, The Ohio State University, Columbus, Ohio 43210, USA}
\affiliation{Center for Cosmology \& Astro-Particle Physics, The Ohio State University, Columbus, Ohio 43210, USA}

\author{Matthew S. B. Coleman}
\affiliation{Research Computing, Princeton University, Princeton, NJ 08544, USA}

\author{Todd A. Thompson}
\affiliation{Center for Cosmology \& Astro-Particle Physics, The Ohio State University, Columbus, Ohio 43210, USA}
\affiliation{Department of Astronomy, The Ohio State University, Columbus, Ohio 43210, USA}
\affiliation{Department of Physics, The Ohio State University, Columbus, Ohio 43210, USA}

\begin{abstract}
The neutrino-driven wind cooling phase of proto-neutron stars (PNSs) follows successful supernovae. Wind models without magnetic fields or rotation fail to achieve the necessary conditions for production of the third $r-$process peak, but robustly produce a weak $r-$process in neutron-rich winds. Using 2D magnetohydrodynamic simulations with magnetar-strength magnetic fields and rotation, we show that the PNS rotation rate significantly affects the thermodynamic conditions of the wind. We show that high entropy material is quasi-periodically ejected from the closed zone of the PNS magnetosphere with the required thermodynamic conditions to produce heavy elements. We show that maximum entropy $S$ of the material ejected depends systematically on the magnetar spin period $P_{\star}$ and scales as $S \propto P_{\star}^{-5/6}$ for sufficiently rapid rotation. We present results from simulations at a constant neutrino luminosity representative of $\sim 1-2$\,s after the onset of cooling for $P_{\star}$ ranging from 5\,ms to 200\,ms and a few simulations with evolving neutrino luminosity where we follow the evolution of the magnetar wind until $10-14$\,s after the onset of cooling. We estimate at magnetar polar magnetic field strength $B_0=3\times 10^{15}$\,G and $10^{15}$\,G that neutron-rich magnetar winds can respectively produce at least $\sim 1-5\times 10^{-5}$\,M$_{\odot}$ and $\sim 1-4\times 10^{-7}$\,M$_{\odot}$ of material with the required parameters for synthesis of the third $r-$process peak, within $1-2$\,s and 10\,s respectively in that order after the onset of cooling. We show that proton-rich magnetar winds can have favorable conditions for production of $p-$nuclei, even at a modest $B_0=5\times 10^{14}$\,G.    
\end{abstract}

\section{Introduction}
\label{intro}
Core collapse of massive stars with mass $\gtrsim 8-10$\,M$_{\odot}$ and subsequent successful explosion results in the formation of proto-neutron stars (PNSs). PNSs cool by emitting neutrinos \citep{Burrows1986}, which can drive a thermal wind \citep{Duncan1986, Burrows1995, Janka1996}. PNS winds have been suggested as promising astrophysical sites for production of heavy elements \citep{Woosley1992, Meyer1992, Woosley1994}, but earlier wind models do not achieve the required conditions for production of the third $r-$process peak \citep{QW1996}, except for a very compact PNS at high neutrino luminosities that occur $\lesssim 1$\,s after the onset of the cooling phase \citep{Otsuki2000, Thompson2001, Wanajo2001}.

Magnetar-strength magnetic fields $\gtrsim 10^{14}-10^{15}$\,G may be important for nucleosynthesis in PNS winds. One possible explanation for the origin of strong magnetic fields in PNSs is rapid rotation with periods $\lesssim 3$\,ms \citep{Duncan1992, Thompson1993}. But, the fact that magnetars are common in the Galaxy comprising $\sim 40\%$ of all neutron star births \citep{Beniamini2019}, and a lack of evidence for anomalously energetic supernova remnants associated with anomalous X-ray pulsars and soft gamma-ray repeaters \citep{Vink2006, Martin2014}, suggests that rapid rotation of the order of milliseconds may not be required to generate magnetar-strength magnetic fields. \cite{White2022} find that even modest progenitor rotation may be sufficient to produce magnetars during core-collapse (see also \citealt{Thompson1993}). Other mechanisms for generating magnetar strength magnetic fields such as strong fossil fields from the progenitor star (e.g., \citealt{Ferrario2006}) and fallback accretion (e.g., \citealt{Barrere2022}) have been suggested.  

\cite{Thompson2003} show that magnetar-strength magnetic fields can trap matter in the closed zone of the magnetosphere and that energy deposition by neutrinos can lead to ejection of high entropy material. \cite{Thompson2003} also provide the first estimates of entropy enhancement as a function of magnetic field strength and predict a robust $r-$process in magnetar winds. \cite{Thompson2018} perform the first 2D magnetohydrodynamic (MHD) simulations to show entropy enhancement in dynamical eruptions (plasmoids) from the PNS magnetosphere for a non-rotating PNS. \cite{Desai2023} study the dynamics of magnetized non-rotating PNS winds in 3 dimensions and show entropy enhancement in plasmoids.  

The electron fraction $Y_{\rm e}$ of the wind is one of the major parameters that affects nucleosynthesis. $Y_{\rm e}$ in the wind is mostly set by luminosities and energies of the electron-type neutrinos and antineutrinos via the charged current processes \citep{QW1996}. $Y_{\rm e}$ thus depends on the PNS cooling model (e.g., \citealt{Pons1999, Vartanyan2023}). However, \cite{Metzger2008} show that magnetocentrifugal slinging caused by very rapid rotation (near-breakup) and strong magnetic fields can lead to a decrease in asymptotic $Y_{\rm e}$. Some studies suggest that PNS winds may be moderately neutron-rich and thus facilitate $r-$process, but possibly proton-rich at late times during the cooling phase (e.g., \citealt{Roberts2012, Roberts2012_2}).

The synthesis of $p-$nuclei in proton-rich winds has been previously studied \citep{Hoffman1996, Pruet2005, Pruet2006, Frohlich2006, Wanajo2006, Wanajo2011}. \cite{Pruet2006} show that the wind entropy has a significant impact on nucleosynthesis. By artificially increasing the entropy by a factor of $2-3$ over the baseline entropy of $55-77$\,$\rm k_B \ baryon^{-1}$, they find that the nucleosynthesis can extend until $A\approx 170$. \cite{Friedland2023} show that for PNS masses $\gtrsim 1.7$\,M$_{\odot}$, the $\nu p-$process can synthesize $p-$nuclei until $A\lesssim 105$ in neutrino-driven outflows of core collapse supernovae. The elemental abundances between Strontium and Silver in very metal poor stars may be a result of weak $r-$process and $\nu p-$process occurring in core-collapse supernova explosions \citep{Psaltis2023}. \cite{Xiong2023} suggest a new nucleosynthesis process ($\nu r$-process) for the production of $p-$nuclei in neutron-rich winds which occurs if the product of expansion timescale and the rate for $\nu_{\rm e}$ absorption on neutrons exceeds 0.1 at the $T\sim3\times 10^{9}$\,K (3\,GK) surface.

An important piece of physics not included in earlier works by \cite{Thompson2018} and \cite{Desai2023} is magnetar rotation. In our earlier works, we explored 2D MHD simulations of magnetar winds with rotation and magnetar-strength magnetic fields in the context of magnetar spindown \citep{Prasanna2022}, gamma-ray bursts and superluminous supernovae \citep{Prasanna2023}. \cite{Metzger2007} use 1D models with magnetic fields and rotation to argue that rapidly rotating magnetars produce comparatively more favorable conditions for $r-$process than slowly rotating non-magnetized PNSs, but due to a lack of dynamic PNS magnetosphere in their models, they do not explore the thermodynamics of plasmoids. \cite{Vlasov2014, Vlasov2017} explore the prospects for nucleosynthesis in magnetar winds including the effects of rotation, but assume a static magnetosphere which precludes the study of plasmoid dynamics.  

In this paper, we perform the first 2D MHD simulations of magnetar winds with magnetar-strength magnetic fields and rotation to study the prospects for nucleosynthesis. We show that the PNS rotation rate significantly affects the entropy of the material in plasmoids. Entropy is a critical parameter that influences nucleosynthesis in both neutron-rich (e.g., \citealt{Thompson2018}) and proton-rich (e.g., \citealt{Pruet2006}) winds. We show that magnetar winds can have favorable conditions for the production of $r-$process nuclei, $p-$nuclei, and the $\nu r-$process.  

In Section \ref{section:model}, we describe the numerical method, microphysics, and boundary conditions in the simulations. In Section \ref{results}, we present the results to show that PNS rotation rate significantly influences the plasmoid entropy and other wind properties. In Section \ref{conclusion}, we summarize the results and discuss the implications of our results for nucleosynthesis in magnetar winds.

\section{Model}
\label{section:model}
We use the MHD code Athena\texttt{++} \citep{Stone2020} for our simulations, which we have configured to solve the non-relativistic MHD equations:
\begin{gather}
          \frac{\partial \rho}{\partial t} + \nabla\cdot\left(\rho\bm{v}\right)=0,\label{eq:continuity}\\
          \frac{\partial \left(\rho\bm{v}\right)}{\partial t} + \nabla\cdot\left[\rho\bm{vv}+\left(P+\dfrac{B^2}{2}\right)\mathbf{I}-\bm{BB}\right]=-\rho \frac{GM_{\star}}{r^2}\boldsymbol{\hat{r}},\label{eq:momentum}\\
          \frac{\partial E}{\partial t} + \nabla\cdot\left[\left(E+\left(P+\dfrac{B^2}{2}\right)\right)\bm{v}-\bm{B}\left(\bm{B}\cdot\bm{v}\right)\right]=\dot{Q},\label{eq:energy}\\
          \frac{\partial \bm{B}}{\partial t} -\nabla\times\left(\bm{v}\times\bm{B}\right)=0,\label{eq:eulermag}
\end{gather}
where $M_{\star}$ is the mass of the PNS, $r$ is the radius from the center of the PNS, $\rho$ is the mass density of the fluid, $\bm{v}$ is the fluid velocity, $E$ is the total energy density of the fluid, $P$ is the fluid pressure, $\dot{Q}$ is the neutrino heating/cooling rate, and $\bm{B}$ is the magnetic field. We solve the MHD equations in two dimensions using spherical polar coordinates assuming axisymmetry. We assume that the magnetic and rotation axes of the PNS are aligned. 

\subsection{Microphysics}
\label{micro}
We use the general equation of state (EOS) module in Athena\texttt{++} \citep{Coleman2020}\footnote{With additional modifications for a composition-dependent EOS.} and run the simulations using the approximate analytic form of the general EOS \citep{QW1996} containing non-relativistic baryons, relativistic electrons and positrons, and photons. This is a reasonable approximation to the EOS at temperatures $T \gtrsim 0.5$\,MeV, but this approximation is not valid at temperatures much lower than 0.5\,MeV because the electrons and positrons cease to be relativistic. We use the approximate analytic EOS to limit the computation time. We have run a few simulations with the general Helmholtz EOS \citep{Timmes2000}, which is accurate even in the regions with $T<0.5$\,MeV, to compare with the results obtained using the approximate analytic EOS. We find reasonable agreement between the two at $T\gtrsim 0.5$\,MeV (see Table \ref{table1}). 

We evolve the electron fraction $Y_{\rm e}$ of the wind as a function of radius and time as described in \cite{Prasanna2023}. To compute the neutrino energy deposition rate $\dot{q}$ (related to $\dot{Q}$ in equation \ref{eq:energy} as $\dot{Q}=\dot{q}\rho$), we consider the neutrino heating and cooling rates resulting from charged current $\nu_{\rm e}$ and $\bar{\nu}_{\rm e}$ absorption processes, neutrino scattering on electrons and positrons, neutrino scattering on nucleons, and neutrino-antineutrino annihilation to form electron-positron pairs. We use the expressions from \cite{Thompson2001}, but we ignore the general relativistic (GR) terms. GR effects increase the effective gravitational potential, and the associated gravitational redshift terms decrease the heating rate, both leading to lower $\dot{M}$, larger entropy, and shorter expansion timescales (\citealt{Cardall1997}; geodesic bending partially compensates the gravitational redshift terms \citep{Salmonson1999}). For comparison, we present results from two simulations with only the charged current neutrino heating and cooling processes (see Table \ref{table1}) and find comparable results.   

\subsection{Reference frame, initial conditions, and computation grid}
\label{ICs}
We perform the simulations in an inertial (lab) frame of reference. We initialize the simulations with a one dimensional non-rotating and non-magnetic (NRNM) wind model. We incorporate an initial purely dipolar magnetic field using the magnetic vector potential (see \citealt{Prasanna2022} for details). To include the effects of rotation in the initial conditions, we initialize the azimuthal velocity $v_{\phi}$ by conserving the specific angular momentum: $v_{\phi}(r)=r\Omega_{\star}\sin \theta \left(R_{\star}/r\right)^2$, where $r$ is the radius from the center of the PNS, $\theta$ is the polar angle measured from the rotation axis of the PNS, $R_{\star}$ is the PNS radius (12\,km in our simulations), and $\Omega_{\star}=2\pi/P_{\star}$ is the angular frequency of rotation of the PNS. The computation grid starts from the surface of the PNS and extends to an outer boundary radius of $10^{4}$\,km. The grid is divided into $N_r$ logarithmically spaced zones in the radial direction and $N_{\theta}$ uniformly spaced zones  in the $\theta$ direction. For the simulations in this paper, we use 512 radial zones and 256 $\theta$ zones. We present results from a simulation with a higher resolution containing 1024 radial zones and 512 $\theta$ zones and one simulation with a low resolution containing 256 radial zones and 128 $\theta$ zones for comparison (see Table \ref{table1}). We find comparable results with all the three choices, but we stick with ($N_r,\ N_{\theta}$)=($512,\ 256$) as a compromise between computational complexity and sufficient resolution.

\subsection{Boundary conditions}
\label{BCs}
We enforce co-rotation with the PNS and set $v_{\phi}(r)=r\Omega_{\star}\sin\theta$ at the inner boundary. We solve equation \ref{eq:momentum} to obtain the inner boundary conditions (BCs) on density, radial velocity $v_r$, and $\theta-$velocity $v_{\theta}$. We set $v_r=0$ and $v_{\theta}=0$ at the inner boundary. We set $v_r=0$ at the inner boundary to ensure that the $\phi-$derivative of $\rho$ is zero (axisymmetry condition, see \citealt{Prasanna2022}), although it is incompatible with a steady state wind. To test if this condition affects the results, we have run one simulation with $v_r$ at the inner boundary set by conserving the mass outflow rate $\dot{M}$ (equation \ref{Mdot}). We find good agreement between the two choices for the $v_r$ inner BC (see Table \ref{table1}).

The density inner BC is (refer to \citealt{Prasanna2022} for the derivation):
\begin{equation}\label{densBC}
\begin{split}
\rho(r,\theta)=\rho_0\exp\left[\frac{GM_{\star}m_{\rm n}}{k_{\rm B}T} \left(\frac{1}{r} -\frac{1}{R_{\star}}\right)\right] \\
    \times \exp\left[\frac{m_{\rm n}r^2\Omega_{\star}^2\sin^2\theta }{2k_{\rm B}T}\right] \exp \left[\frac{-m_{\rm n} R_{\star}^2\Omega_{\star}^2}{2k_{\rm B}T}\right],
    \end{split}
\end{equation}
where we have assumed that the pressure near the surface is dominated by ideal nucleon gas pressure. In equation \ref{densBC}, $m_{\rm n}$ is the average mass of a nucleon, $\rho_0$ is the base density at the equator on the PNS surface, $T$ is the temperature, and $k_{\rm B}$ is the Boltzmann constant. We set $\rho_0=10^{12}$\,g cm$^{-3}$, which is a reasonable choice for base density at the surface of a PNS (see \citealt{Thompson2001}, where the density boundary condition is set using the neutrino optical depth). We have tested our 1D NRNM simulations with base densities $10^{11}-10^{13}$\,g cm$^{-3}$ and find that the mass outflow rate is not very sensitive to the base density across this range. For comparison with the results obtained using the density BC in equation \ref{densBC}, we present results from simulations with a constant density inner BC (see Table \ref{table1}). We find that results obtained from both the choices of density BCs agree with each other. However, we stick with equation \ref{densBC} for the inner density BC because this condition satisfies equation \ref{eq:momentum}.      

We set the azimuthal component of the electric field $E_{\phi}$ to zero at the inner edge of the first active computational zone to maintain a constant magnetic field at the surface of the PNS as a function time \citep{Prasanna2023}. At the PNS surface, kinetic equilibrium is established between the neutrino flux and the wind due to high temperature and density \citep{Burrows1982, QW1996}. Hence, we enforce net zero neutrino heating ($\dot{q}=0$) at the inner boundary. To set the temperature $T$ and electron fraction $Y_{\rm e}$ in the ghost zones at the inner boundary, we solve the simultaneous equations $\dot{q}=0$ and $\dot{Y}_{\rm e}=0$ (equation 7 in \citealt{QW1996}) using the two dimensional Newton-Raphson (2DNR) method at the first radial ghost zone for each value of the polar angle $\theta$. We use the value of $T$ obtained by this method to set the temperature in the other radial ghost zones at a given $\theta$ at the inner boundary (constant temperature condition in the radial direction). Once $T$ is set in all the ghost zones, we compute the electron fraction $Y_{\rm e}$ in the ghost zones by solving $\dot{q}=0$ using the one dimensional Newton-Raphson method. We name this the ``first choice inner BCs'' for ease of referencing. We note that we ignore temperature gradients in both the radial and $\theta$ directions in deriving the density BC (equation \ref{densBC}). But with the ``first choice BCs'', there is a temperature gradient in the $\theta$ direction at the inner boundary. To test if this affects the results, we have run one simulation where the 2DNR procedure is performed at only one ghost zone (the one closest to the surface at the north pole) and the obtained temperature is used to set the value of $T$ in all the other ghost zones. We find similar results with both these BC choices (see Table \ref{table1}). We note that gradients in the $\theta$ direction become important only when $R_{\star}^2\Omega_{\star}^2 \sim k_{\rm B}T/m_{\rm n}$ at the PNS surface, which occurs for $\Omega_{\star}\gtrsim 2000$\,rad s$^{-1}$ ($P_{\star}\lesssim 3$\,ms).

We find that there is no solution to the 2DNR for arbitrary values of base density $\rho_0$ at given neutrino luminosities and energies. There is a maximum value of $\rho_0$ above which there is no simultaneous solution to the set of equations $\dot{q}=0$ and $\dot{Y}_{\rm e}=0$. We briefly explain the reason for this finding. The charged current cooling term (see equation 20 in \citealt{Thompson2001}) is sensitive to the electron degeneracy parameter $\eta$ which in turn depends on the product of density $\rho$ and electron fraction $Y_{\rm e}$ (see equation 6 in \citealt{QW1996}). The charged current heating term is negative near the surface of the PNS which is balanced by the heating caused by neutrino-nucleon interaction (see equation 28 in \citealt{Thompson2001}; we note that there is a minor typo in the last term where $\eta_{e}$ has to be replaced with $\eta_{\nu}$) to produce net zero neutrino heating at the PNS surface (see Figure 6 in \citealt{Thompson2001}). As the value of $\rho_0$ increases at fixed values of neutrino luminosities and energies, we find that the the values of $Y_{\rm e}$ and the temperature $T$ decrease in order to simultaneously satisfy $\dot{q}=0$ and $\dot{Y}_{\rm e}=0$. At a certain maximum value of $\rho_0$ depending on the neutrino luminosities and energies, we find that no further decrease in $T$ and $Y_{\rm e}$ can satisfy the 2DNR equations.     

Another reasonable choice for the inner BCs (we name this the ``second choice BCs'') is to simultaneously enforce $\dot{q}=0$ and $\dot{Y}_{\rm e}=0$ in all the ghost zones. With this choice, the temperature is not the same in all the ghost zones even in the radial direction. But, we note that equation \ref{densBC} still holds for the density inner BC because the temperature near the PNS surface is sufficiently slowly varying such that $|\nabla T| \ll |T \, \nabla \rho/\rho|$ in the radial direction, which allows us to ignore temperature gradients when deriving equation \ref{densBC} for the rotation rates considered in this paper. Due to the nature of the density in the ghost zones at the inner boundary (see equation \ref{densBC}) and the fact that there is no simultaneous solution to $\dot{q}=0$ and $\dot{Y}_{\rm e}=0$ above a certain value of $\rho_0$, this choice of the inner BCs precludes values of $\rho_0 \gtrsim 10^{12}$\,g cm$^{-3}$ for the range of neutrino luminosities considered in this paper (see Section \ref{const_lum_sub} and \ref{evol_sub}). With this choice of BCs, we have to restrict $\rho_0$ to values $\lesssim 5\times 10^{11}$\,g cm$^{-3}$, which is still a reasonable choice for the PNS base density. But as the neutrino luminosity decreases, the density gradient near the PNS surface increases and this leads to a drop in the density in the first active computational zone. In order to maintain a density $\geq 3\times 10^{11}$\,g cm$^{-3}$ in the first active zone in our evolutionary models (see Section \ref{evol_sub}), we choose to perform the simulations with the first choice inner BCs (as described in the previous paragraph). With the first choice, we find that the neutrino luminosities in the evolutionary models can reach $10^{51}$\,ergs s$^{-1}$, which is sufficient for our purposes in this paper. For comparison, we present some results with the second choice inner BCs and $\rho_0=3\times 10^{11}$\,g cm$^{-3}$ (see Table \ref{table1}). We find good agreement between the results obtained from both these choices.             

At the outer boundary, we set outflow conditions by conserving the specific angular momentum and the mass outflow rate $\dot{M}$ (equation \ref{Mdot}). 

\section{Results}
\label{results}
\subsection{Diagnostic quantities}
\label{diag}
The entropy $S$ in the wind, dynamical expansion timescale $t_{\rm exp}$, the figure of merit parameter for $r-$process nucleosynthesis $\zeta$, mass outflow rate $\dot{M}$, angular momentum flux $\dot{J}$, spindown timescale $\tau_{\rm J}$, and energy outflow rate $\dot{E}$ are the principal physical quantities we measure. The dynamical expansion timescale is defined as:
\begin{equation}
\label{texp}
    t_{\rm exp}=\frac{1}{v_r}\left|\frac{1}{T}\frac{dT}{dr}\right|^{-1}.
\end{equation}
The figure of merit parameter for $r-$process nucleosynthesis $\zeta$ at $T\sim 0.5$\,MeV is given by \citep{Hoffman1997},
\begin{equation}
\label{zeta_eqn}
    \zeta=\frac{S^3}{Y_{\rm e}^3 \left(1.28 \ t_{\rm exp}\right)}.
\end{equation}
The critical value of $\zeta$ required for production of the third $r-$process peak is $\zeta_{\rm crit}\simeq 8\times 10^{9} \ ({\rm k_{B} \ baryon^{-1}})^3 \ {\rm s}^{-1}$ \citep{Hoffman1997} at $T\sim 0.5$\,MeV. The mass outflow rate through a closed spherical surface is given by,
\begin{equation}
\label{Mdot}
    \dot{M} \left(r \right)= \oint_S r^2 \rho v_r d\Omega,
\end{equation}
where $d\Omega$ is the solid angle. The $z$-component (along the axis of rotation of the PNS) of the angular momentum flux is given by the following integral over a closed spherical surface \citep{Vidotto2014}:
\begin{equation}
\label{Jdot}
    \dot{J}\left(r \right)= \oint_S \left[-\frac{B_rB_{\phi}r\sin \theta}{4\pi}+\rho v_r v_{\phi}r\sin \theta\right]r^2 d\Omega.
\end{equation}
In a time-steady state, $\dot{M}\left(r \right)$ and $\dot{J}\left(r \right)$ are constant as a function of radius (except in the first $\sim 10$ zones near the inner boundary where boundary effects affect the profiles).

The total angular momentum of the star is roughly $J=\frac{2}{5}MR_{\star}^2\Omega_{\star}$. We define the spindown time of the PNS as:
\begin{equation}
\label{tauj}
    \tau_{\rm J}=\frac{J}{\dot{J}}.
\end{equation}
Since $\dot{J}\left(r \right)$ is constant with radius, $\tau_{\rm J}$ can be measured at any radius (see Section \ref{const_lum_sub} for details). 

The energy flux is given by the following surface integral (we generalize the definition in \citealt{Metzger2007} to two and three dimensions):
\begin{align}
\label{Edot}
    \dot{E}\left(r \right)&= \oint_S r^2 \rho v_r\left[\frac{1}{2}\left(v_r^2+v_{\theta}^2+v_{\phi}^2\right) -\frac{rB_rB_{\phi}\Omega_{\star}\sin\theta}{4 \pi \rho v_r}\right.\\[-20pt]\nonumber &\qquad -\left.\frac{GM_{\star}}{r} +e+\frac{P}{\rho}\right] d\Omega,
\end{align}
where $e$ is the specific internal energy of the outflow.

\subsection{Plasmoids from the PNS magnetosphere}
\label{plasmoid_sub}
\begin{figure*}
\centering{}
\includegraphics[width=\textwidth]{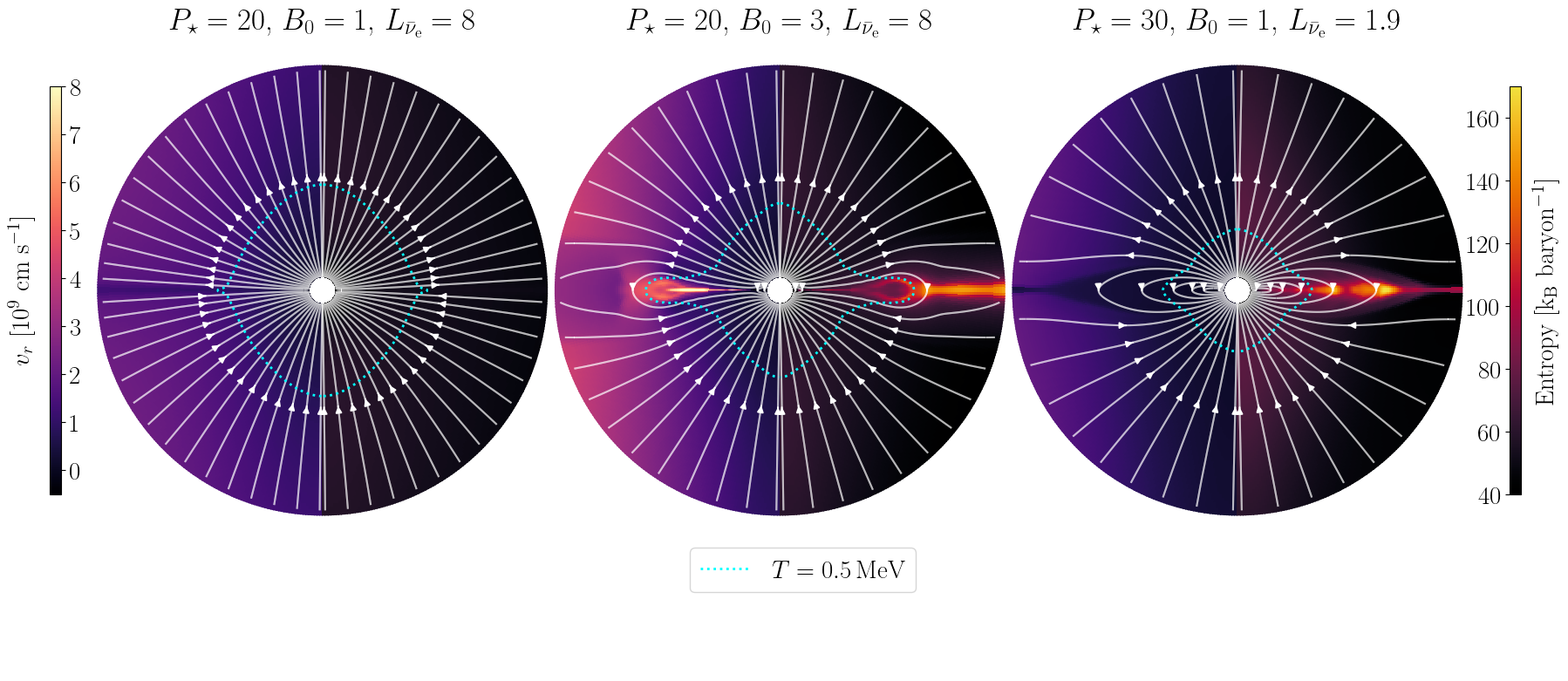}
\vspace{-14mm}
\caption{Comparison of PNS magnetosphere for different values of polar magnetic field strength $B_0$ (units of $10^{15}$\,G) and neutrino luminosity (units of $10^{51}$\,ergs s$^{-1}$). The spin period $P_{\star}$ is in units of milliseconds. As the magnetic field strength dominates the wind pressure gradient (see Section \ref{plasmoid_sub} for details), we find quasi-periodic dynamical plasmoid eruptions from the closed zones of the PNS magnetosphere as shown in the middle and right panels, otherwise the magnetic field settles into a stable split-monopole field configuration as shown in the left panel. This figure shows 2D maps of radial velocity $v_r$ (left half of each panel) and entropy (right half of each panel). The minimum value of entropy shown on the colorbar is chosen for better contrast, the actual minimum value of entropy in the simulations is smaller. The central white circle is the PNS with a radius of 12\,km and the white lines are the magnetic field lines. The magnetic and rotation axes of the PNS are along the vertical in the figure. The 0.5\,MeV temperature surface is shown by the cyan dotted lines. The outer boundary in this figure is at a radius of 200\,km. The left and the right panels show data from an evolutionary model (see Section \ref{evol_sub}), and the middle panel shows data from a fixed neutrino luminosity model (see Section \ref{const_lum_sub}). Although the PNS spin periods in the panels are slightly different, they do not significantly affect the structure of the magnetosphere. The structure of the magnetosphere in this figure is determined mostly by $B_0$ and the neutrino luminosity.} 
\label{plasmoids}
\end{figure*}

As predicted by \cite{Thompson2003} and shown by \cite{Thompson2018}, we find quasi-periodic eruptions from the PNS magnetosphere when the magnetic field is strong enough. Figure \ref{plasmoids} shows 2D maps of radial velocity $v_r$ and entropy along with the structure of the PNS magnetosphere for different values of magnetic field strength and neutrino luminosity. If the magnetic tension force in the PNS equatorial region cannot exceed the wind pressure gradient, the wind breaks open the closed zone of the magnetic field and the magnetic field structure settles into a stable split-monopole configuration as shown in the left panel of Figure \ref{plasmoids}. If the magnetic tension force is sufficiently strong compared to the wind pressure gradient, the closed zone of the magnetic field in the equatorial region traps matter in a helmet streamer configuration, as in models of Solar Corona (e.g., \citealt{Pneuman1971, Steinolfson1982}). The closed magnetosphere is unstable because the neutrinos from the PNS deposit energy and heat the trapped matter. Eventually, when the wind pressure gradient exceeds the magnetic tension force, the trapped matter is ejected in a toroidal plasmoid. The magnetic field quickly reconnects and the process repeats again. Plasmoids erupt quasi-periodically and the ejection timescale depends on the magnetic field strength ($t_{\rm ej}\propto B^2$, see \citealt{Thompson2018} for analytic estimates in the non-rotating case). We show in this paper that the ejection timescale and properties of the plasmoids also depend on the PNS spin period. 

The magnetic tension force can exceed the wind pressure gradient in two ways: a larger value of polar magnetic field strength $B_0$ at a fixed neutrino luminosity (middle panel in Figure \ref{plasmoids}) or a smaller value of neutrino luminosity at a fixed value of $B_0$ (right panel in Figure \ref{plasmoids}). In Figure \ref{plasmoids}, the left and the right panels show data from an evolutionary model (starting from an initial spin period of 20\,ms at a fixed $B_0=10^{15}$\,G, see Section \ref{evol_sub}), and the middle panel shows data from a fixed neutrino luminosity model (see Section \ref{const_lum_sub}).

\begin{figure*}
\centering{}
\includegraphics[width=\textwidth]{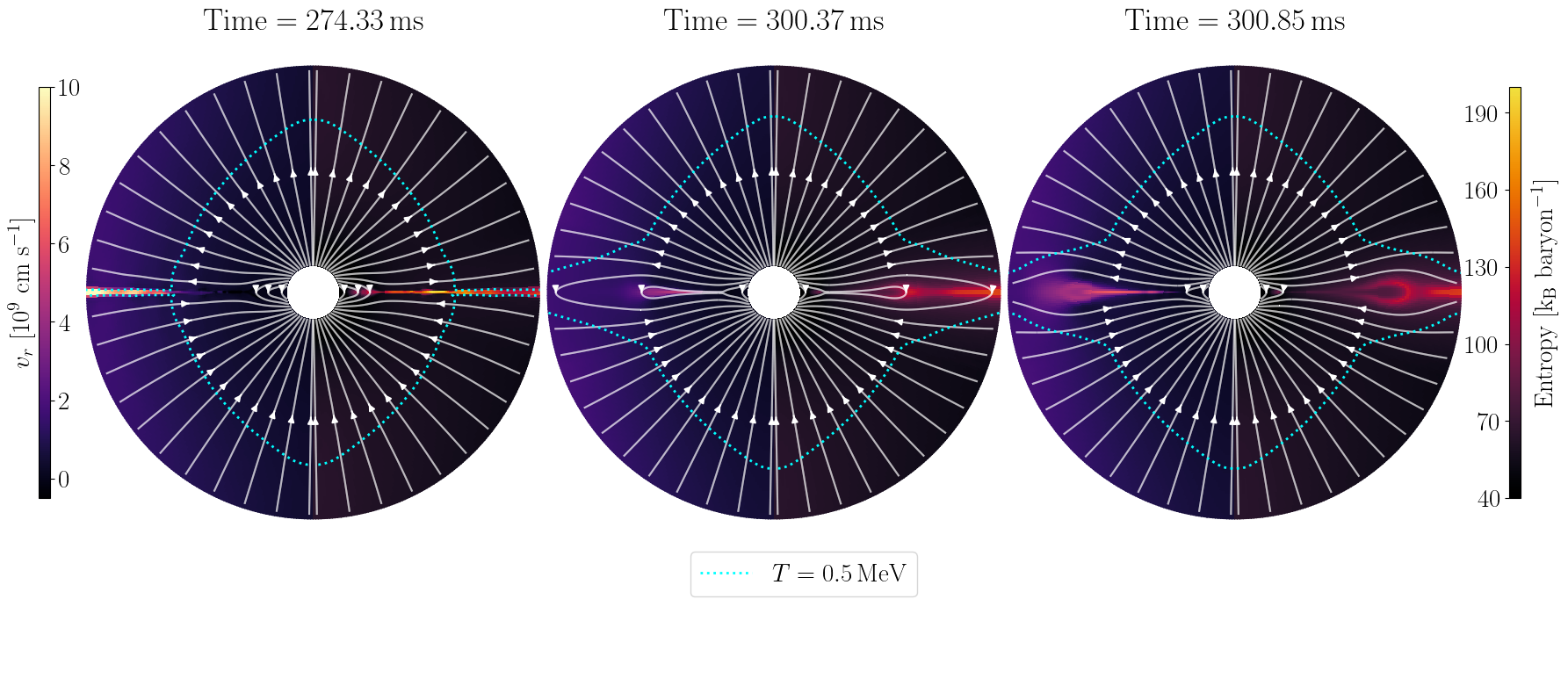}
\vspace{-14mm}
\caption{Sequence of images showing the structure of the magnetic field during a plasmoid eruption. This figure shows 2D maps of radial velocity $v_r$ and entropy as a function of time from the start of the simulation at a PNS spin period of 20\,ms and polar magnetic field strength $B_0=3\times 10^{15}$\,G from a constant neutrino luminosity simulation (see Section \ref{const_lum_sub}). The minimum value of entropy shown on the colorbar is chosen for better contrast, the actual minimum value of entropy in the simulations is smaller. The high entropy and $v_r$ regions in the left panel are from a previous plasmoid eruption. The outer boundary in this figure is at a radius of 100\,km. The 0.5\,MeV temperature surface is shown by the cyan dotted lines. The white lines are the magnetic field lines and the central white circle is the PNS with a radius of 12\,km. } 
\label{plasmoids_sequence}
\end{figure*}

Figure \ref{plasmoids_sequence} is a sequence of images showing the structure of the magnetic field during a plasmoid eruption along with 2D maps of radial velocity $v_r$ and entropy as a function of time from the start of the simulation, at a constant PNS spin period of 20\,ms and polar magnetic field strength $B_0=3\times 10^{15}$\,G from a constant neutrino luminosity simulation (see Section \ref{const_lum_sub}). As shown in the figure, the first closed magnetic field line in the equatorial region remains unchanged at different times because the magnetic tension force at this field line is so large that the the wind pressure gradient can never exceed the magnetic tension. The second field line in the equatorial region shows the magnetic reconnection event and the eruption of matter. 

\subsection{Constant neutrino luminosity simulations}
\label{const_lum_sub}
We first present results from simulations with constant neutrino luminosities and energies at a polar magnetic field strength $B_0=3\times10^{15}$\,G. For these simulations, the neutrino luminosities are $L_{\rm \bar{\nu}_e}=8\times10^{51}$\,ergs s$^{-1}$, $L_{\rm \nu_e}=L_{\rm \bar{\nu}_e}/1.3$, and $L_{\rm \nu_{\mu}}=L_{\rm \bar{\nu}_{\mu}}=L_{\rm \nu_{\tau}}=L_{\rm \bar{\nu}_{\tau}}=L_{\rm \bar{\nu}_e}/1.4$, and the neutrino energies are $\epsilon_{\rm \bar{\nu}_e}=14$\,MeV, $\epsilon_{\rm \nu_e}=11$\,MeV, and $\epsilon_{\rm \nu_{\mu}}=\epsilon_{\rm \bar{\nu}_{\mu}}=\epsilon_{\rm \nu_{\tau}}=\epsilon_{\rm \bar{\nu}_{\tau}}=23$\,MeV following \cite{Thompson2001}. All the results presented in this paper are for a $1.4$\,$\rm M_{\odot}$ PNS (all else equal, a higher mass PNS produces higher entropy and lower mass outflow rate \citep{QW1996, Thompson2001}). As described in Section \ref{intro}, the asymptotic electron fraction in the wind is mostly set by neutrino luminosities and energies \citep{QW1996}. For the choices of neutrino luminosities and energies described above, the asymptotic electron fraction is $\sim 0.45$.

\begin{figure*}
\centering
\includegraphics[width=\textwidth]{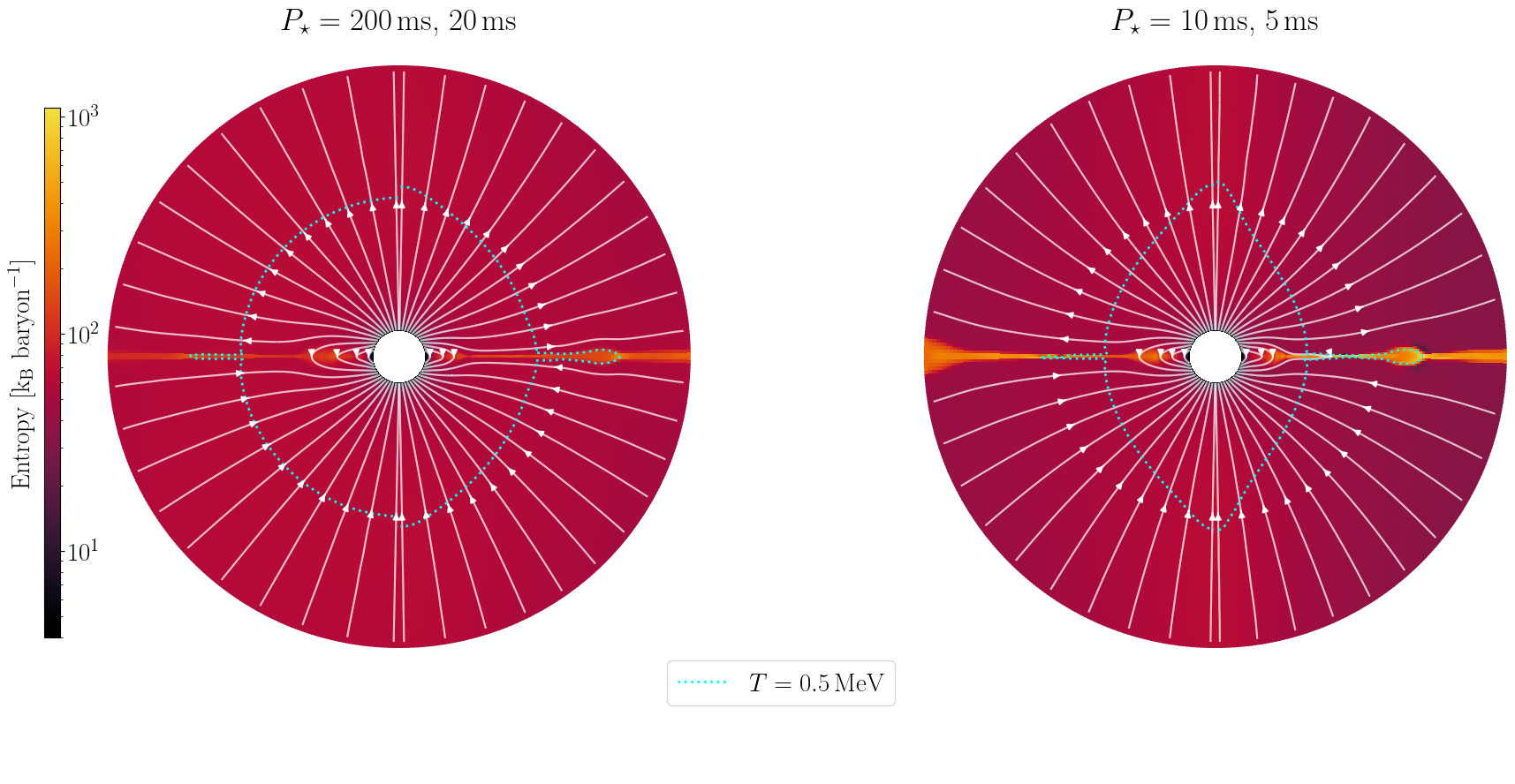}
\vspace{-12mm}
\caption{2D maps of entropy at a polar magnetic field strength $B_0=3\times 10^{15}$\,G for PNS spin periods $P_{\star}=200$\,ms (left half of left panel), 20\,ms (right half of left panel), 10\,ms (left half of right panel), and 5\,ms (right half of right panel) at the time instants of the respective simulations when maximum entropy occurs at the 0.5\,MeV temperature surface. The $T=0.5$\,MeV surface is shown by the cyan dotted lines. The white lines are the magnetic field lines and the central white circle is the PNS with a radius of 12\,km. The outer boundary in this figure is at a radius of 130\,km. These maps are from constant neutrino luminosity simulations (see Section \ref{const_lum_sub}).} 
\label{entr2d}
\end{figure*}

Figure \ref{entr2d} shows the maps of entropy for PNS spin periods $P_{\star}=200$\,ms, 20\,ms, 10\,ms, and 5\,ms at $B_0=3\times 10^{15}$\,G from the constant neutrino luminosity simulations at the time instants of the respective simulations when maximum entropy occurs at the $T=0.5$\,MeV temperature surface. The maps clearly show that the maximum entropy in the equatorial region increases with faster PNS rotation at a fixed value of $B_0$ and neutrino luminosity. We provide the physical intuition for the variation of entropy with PNS rotation rate in Section \ref{scaling_sub}.
 
Table \ref{table1} presents results from the constant neutrino luminosity simulations as a function of PNS spin period $P_{\star}$ at a polar magnetic field strength $B_0=3\times 10^{15}$\,G. We have run these simulations for a duration of 0.5\,s. The quantities presented in Table \ref{table1} are as follows.  $S_{\rm max,0.5}$ is the maximum entropy of the outflowing material at the $T=0.5$\,MeV temperature surface during the course of the simulation. As the magnetic field reconnects following a plasmoid eruption, some wind material is dragged back into the PNS magnetosphere. We do not report the entropy of material with $v_r<0$ that is dragged back. We measure only the entropy of the outflowing material ($v_r>0$) that is released into the expanding wind. $\langle S \rangle _{\dot{M}}$ in Table \ref{table1} is the average entropy of the outflowing material weighted by $\dot{M}$ in the zones within an angle of 3$^{\circ}$ on either side of the equator at the $T=0.5$\,MeV surface. The values reported in the parentheses are the minimum and maximum values of $\langle S \rangle _{\dot{M}}$ during the course of the simulation. We measure the mass flux $\dot{M}_{\zeta, 0.5}$ of the outflowing matter with the $r-$process figure of merit parameter (equation \ref{zeta_eqn}) $\zeta \ge\zeta_{\rm crit}$ through the $T=0.5$\,MeV surface to estimate the nucleosynthetic yields of elements in the third $r-$process peak. We also present the minimum value during the simulation of the average dynamical expansion timescale (equation \ref{texp}) weighted by $\dot{M}$. Although the $T=0.5$\,MeV surface is not spherical, we do not consider the non-radial surface normal to compute $\dot{M}$ through the $T=0.5$\,MeV surface because we sum the mass outflow rate through individual cells of the computational domain ($=r^2\rho v_rd\Omega$, where $d\Omega$ is the solid angle) at $T=0.5$\,MeV and we are interested in only the radial component of the mass flux through the cells which can flow into the expanding wind.

As a sanity check for the measured mass outflow rate of high-$\zeta$ material through the non-spherical $T=0.5$\,MeV surface, we have computed the mass outflow rate $\dot{M}_{\zeta}$ that satisfies the condition $\zeta\ge \zeta_{\rm crit}$ through a spherical surface at a radius of 100\,km. We choose $r=100$\,km because this is the average radius of the $T=0.5$\,MeV surface at the PNS equator in the constant neutrino luminosity simulations. We find that the value of $\dot{M}_{\zeta,0.5}$ through the $T=0.5$\,MeV surface is larger than $\dot{M}_{\zeta}$ through the spherical surface at $r=100$\,km by $43\%$, $3\%$, and $20\%$ for $P_{\star}=200$\,ms, 20\,ms, and 5\,ms respectively, and smaller by $0.2\%$ at $P_{\star}=10$\,ms at $B_0=3\times 10^{15}$\,G in the constant neutrino luminosity simulations. We attribute the difference to the fact that the distance of $T=0.5$\,MeV from the PNS in the equatorial region is modulated by plasmoids and is not always equal to 100\,km. Although we present this sanity check, we emphasize that the value of $\dot{M}_{\zeta,0.5}$ through the $T=0.5$\,MeV surface gives the best estimate of the mass outflow rate of the wind material with favorable conditions for the $r-$process.     

As mentioned earlier, we consider only the wind material with $v_r>0$ for the estimates. This is because we do not know if the wind material with $v_r<0$ that is dragged back during magnetic reconnection is eventually released into the outflowing wind. For example, at $P_{\star}=20$\,ms and $B_0=3\times 10^{15}$\,G in the constant neutrino luminosity simulation, we find that $\sim 2\%$ of the wind material that satisfies the condition $\zeta\ge \zeta_{\rm crit}$ has $v_r<0$. Simulations with Lagrangian tracer particles are required to assess if the material that is dragged back into the magnetosphere is eventually released into the outflow, which will be our focus in a future work.

In Table \ref{table1}, $t_{\rm p}$ is the average time interval between plasmoids. Various regions in the PNS equatorial closed zone erupt quasi-periodically with ejection timescales depending on the value of magnetic field and wind pressure gradient. The average time interval between plasmoids presented in Table \ref{table1} is the average time interval between peaks in the energy outflow rate $\dot{E}$. 
We also present time-averaged values of spindown timescale $\tau_{\rm J}$, mass outflow rate $\dot{M}$, and asymptotic energy outflow rate $\dot{E}$ measured over a spherical surface. These quantities have been averaged over time to account for the modulation due to plasmoids. We measure $\tau_{\rm J}$ and $\dot{M}$ at a radius of 50\,km and asymptotic $\dot{E}$ at a radius of 1000\,km. We note that the time-averaged values of $\tau_{\rm J}$ and $\dot{M}$ are negligibly dependent on the radius of measurement as long as they are not measured within the first $\sim 10$ zones near the PNS surface where boundary effects affect the values. The time-averaged value of asymptotic $\dot{E}$ is also negligibly dependent on the radius of measurement as long as it is measured at radii well past the peak of the net neutrino heating rate (see Figure 4 in \citealt{Prasanna2023} for an example of the profile of net neutrino heating rate including only the charged current processes). The values of $\tau_{\rm J}$, $\dot{M}$, and $\dot{E}$ reported in Table \ref{table1} agree with the results in our previous works \citep{Prasanna2022, Prasanna2023}. 

Since we use the approximate analytic form of the general equation of state \citep{QW1996}, the results are accurate only in the temperature range $T\gtrsim 0.5$\,MeV. In our simulations with the approximate EOS and in a few simulations with the more accurate Helmholtz EOS, we find regions of higher entropy at $T<0.5$\,MeV. $S_{\rm max,[0,2,0.5]}$ shown in Table \ref{table1} is the maximum entropy of the outflowing material in the temperature range $0.2$\,MeV$\le T \le0.5$\,MeV. Since our focus in this paper is to show that the PNS rotation rate affects the entropy and other properties of the wind, the approximate EOS is sufficient. In a work that will follow this paper, we will explore the nucleosynthetic yields throughout the computational grid by using the Helmholtz EOS and Lagrangian tracer particles.           

In the simulation with PNS spin period $P_{\star}=5$\,ms and $B_0=3\times 10^{15}$\,G, we find that the Alfv\'en speed and the fast magnetosonic speed in a few regions on the grid approach or exceed the speed of light. Since our simulations do not include relativistic effects, the result at $P_{\star}=5$\,ms needs to be confirmed by running simulations with relativistic effects. This will be our focus in a future work.

\begin{figure*}
\centering
\includegraphics[width=\textwidth]{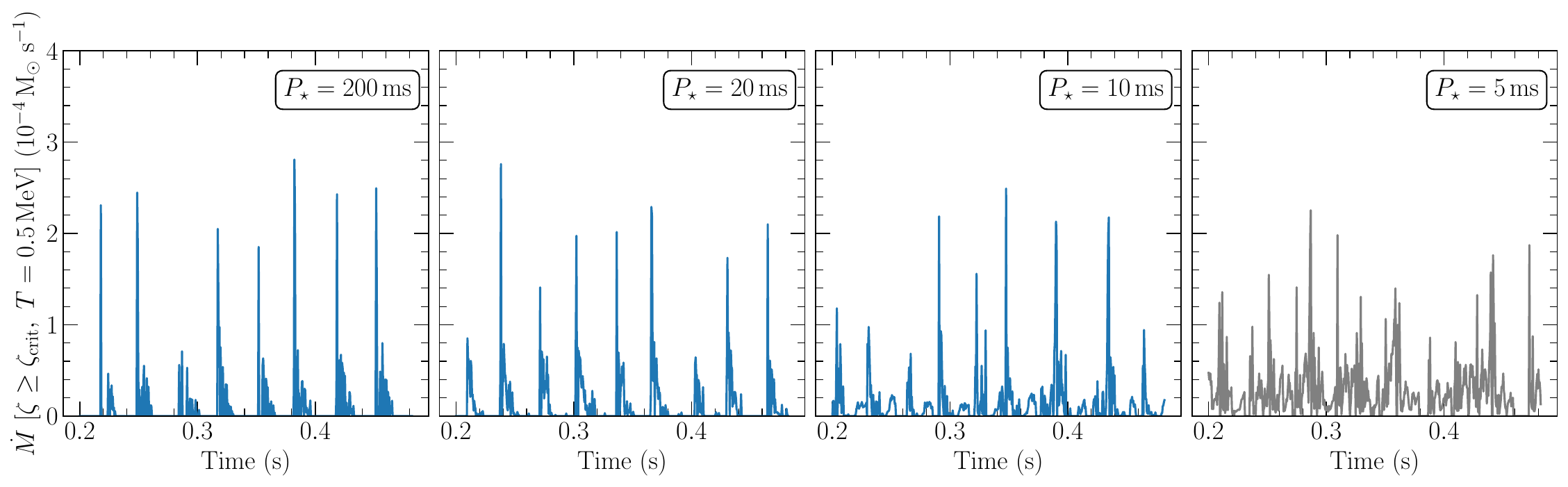}
\caption{Mass outflow rate $\dot{M}$ of the material with the $r-$process figure of merit parameter $\zeta\ge \zeta_{\rm crit}$ (equation \ref{zeta_eqn}) through the $T=0.5$\,MeV surface as a function of time after the start of the simulation for various values of PNS spin period $P_{\star}$ at $B_0=3\times 10^{15}$\,G and constant neutrino luminosity (see Section \ref{const_lum_sub}). Time starts from 0.2\,s on the time axis because we ignore the simulation data that can be affected by the initial condition transient during this time. The profile for $P_{\star}=5$\,ms is shown in gray to emphasize that the Alfv\'en speed and the fast magnetosonic speed exceed the speed of light in this simulation. Inclusion of relativistic effects is required to confirm the result for $P_{\star}=5$\,ms, which will be our focus in a future work. Although there are some regions in the plasmoids where the magnetosonic speeds are superluminal for $P_{\star}=5$\,ms, we do not expect significant qualitative differences in the results pertaining to nucleosynthesis obtained using relativistic and non-relativistic physics. But, we emphasize that relativistic simulations are required to confirm this assertion for $P_{\star}=5$\,ms.} 
\label{mdot_zeta}
\end{figure*}

\begin{figure*}
\centering
\includegraphics[width=\textwidth]{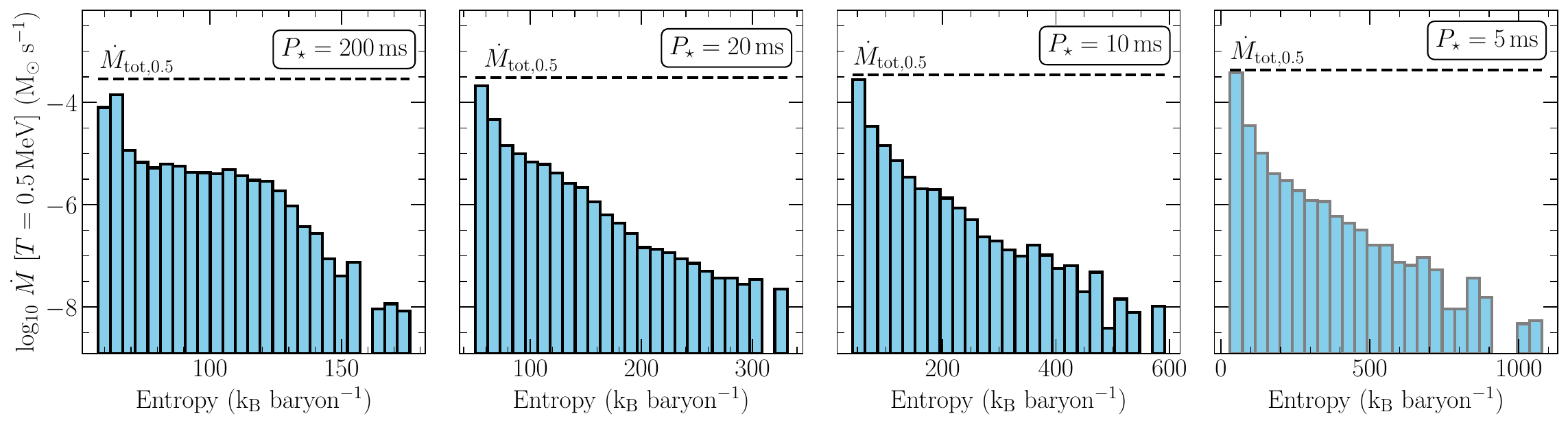}
\caption{Time-averaged mass outflow rate $\dot{M}$ through the $T=0.5$\,MeV surface as a function of entropy for various values of PNS spin period at $B_0=3\times 10^{15}$\,G and constant neutrino luminosity (see Section \ref{const_lum_sub}). The dotted black horizontal line shows the total time-averaged $\dot{M}$ through the $T=0.5$\,MeV surface. The values of $\dot{M}_{\rm tot,0.5}$ in this figure are slightly different from the $\dot{M}$ values computed over a spherical surface reported in Table \ref{table1} because the $T=0.5$\,MeV surface is not spherical (see Figures \ref{plasmoids}, \ref{plasmoids_sequence}, and \ref{entr2d}). The borders in the bar chart for $P_{\star}=5$\,ms are shown in gray to emphasize that the Alfv\'en speed and the fast magnetosonic speed exceed the speed of light in this simulation (refer to the caption of Figure \ref{mdot_zeta}).} 
\label{entr_hist}
\end{figure*}

\begin{figure*}
\centering
\includegraphics[width=\textwidth]{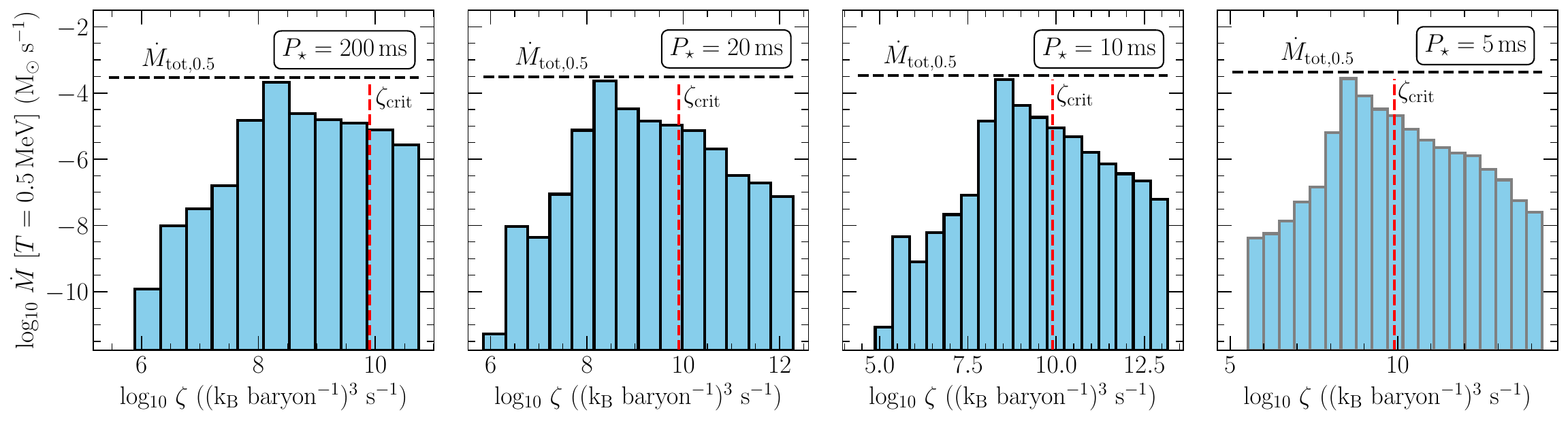}
\caption{Similar to Figure \ref{entr_hist}, but $\dot{M}$ is plotted as a function of $\zeta$ in this figure. The red dotted vertical line shows the value of $\zeta_{\rm crit}$ which estimates the value of $\zeta$ required for the production of the third $r-$process peak \citep{Hoffman1997}.} 
\label{zeta_hist}
\end{figure*}

\begin{figure*}
\centering
\includegraphics[width=\textwidth]{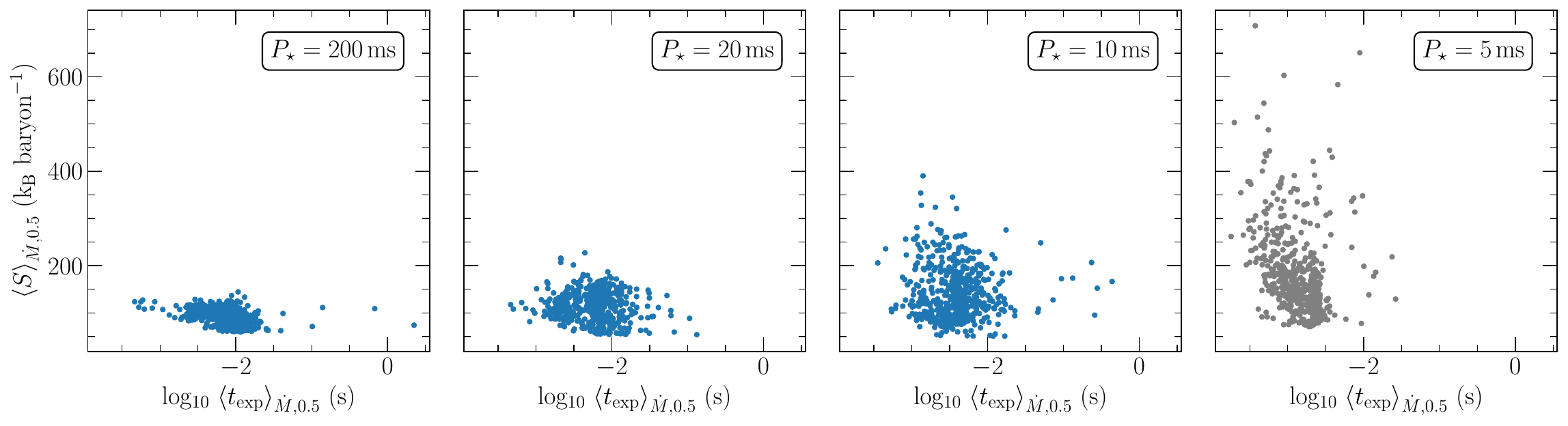}
\caption{Scatter plot of $\dot{M}$ weighted average of entropy and expansion timescale in the zones within 3$^{\circ}$ on either side of the PNS equator at $T=0.5$\,MeV for various values of PNS spin period at $B_0=3\times 10^{15}$\,G. Each dot in the scatter plot represents the weighted average at one time instant. All the dots together in each scatter plot span a duration of 0.3\,s of the simulation time. The data in this figure is from constant neutrino luminosity simulations (see Section \ref{const_lum_sub}). The dots in the plot for $P_{\star}=5$\,ms are shown in gray to emphasize that the Alfv\'en speed and the fast magnetosonic speed exceed the speed of light in this simulation (refer to the caption of Figure \ref{mdot_zeta}).} 
\label{const_lum_scatter}
\end{figure*}

Figure \ref{mdot_zeta} shows the profiles of mass outflow rate $\dot{M}$ through the $T=0.5$\,MeV surface with $\zeta\ge \zeta_{\rm crit}$ (equation \ref{zeta_eqn}) as a function of time after the start of the simulation for different values of PNS spin period at $B_0=3\times 10^{15}$\,G. Figures \ref{entr_hist} and \ref{zeta_hist} show time-averaged $\dot{M}$ through $T=0.5$\,MeV surface as a function of entropy and $\zeta$ respectively. For $P_{\star} \ge 5$\,ms, we find an average mass outflow rate of $\sim 1-3\times 10^{-5}$\,M$_{\odot} \ \rm s^{-1}$ with $\zeta\ge \zeta_{\rm crit}$ through the $T=0.5$\,MeV surface at $B_0=3\times 10^{15}$\,G and neutrino luminosities as described earlier in this subsection. We discuss the implications of this result for nucleosynthesis in Section \ref{conclusion}.

%For comparison, \cite{Thompson2018} find a mass outflow rate of $\sim 8\times 10^{-6}$\,M$_{\odot} \ \rm s^{-1}$ with $\zeta\ge \zeta_{\rm crit}$ at a similar neutrino luminosity and a much higher magnetic field strength of $10^{16}$\,G for a non-rotating PNS (they consider mass in the temperature range 0.1\,MeV\,$\le T \le$\,0.5\,MeV, in contrast to our estimates at $T=0.5$\,MeV only). Although there are differences in microphysics, grid resolution, and boundary conditions, we attribute the smaller mass outflow rate of high-$\zeta$ material in \cite{Thompson2018} to the large magnetic field of $10^{16}$\,G. As the magnetic field strength increases, the time interval between plasmoids increases, resulting in a smaller average mass outflow rate of high-$\zeta$ material.   

Although the average $\dot{M}$ with $\zeta\ge \zeta_{\rm crit}$ is comparable for $P_{\star}\gtrsim 10$\,ms, we find that the thermodynamic conditions of this material are significantly different. Figure \ref{const_lum_scatter} shows a scatter plot of $\dot{M}$ weighted average of entropy and expansion timescale in the zones within 3$^{\circ}$ of the PNS equator at $T=0.5$\,MeV for various values of PNS spin period. As shown in Figures \ref{entr_hist}, \ref{zeta_hist}, and \ref{const_lum_scatter}, the entropy, $\zeta$, and expansion timescale $t_{\rm exp}$ of the material ejected in plasmoids is strongly influenced by the PNS rotation rate. In a future work, we intend to incorporate Lagrangian tracer particles in the simulations to track nucleosynthetic yields to understand how the different thermodynamic conditions affect the elemental yields.

\begin{deluxetable*}{ccccccccccc}
\tablecolumns{11}
\label{table1}
\tablewidth{0pt}
\tablecaption{Wind properties as a function of PNS spin period $P_{\star}$ from constant neutrino luminosity simulations at a polar magnetic field strength $B_0=3\times 10^{15}$\,G (see Section \ref{const_lum_sub}) and from the evolutionary models (see Section \ref{evol_sub}) at $B_0=10^{15}$\,G (run until 10\,s after the onset of cooling).}
\tablehead{
    \colhead{\tablenotemark{\rm \scriptsize a}$B_0$} & \colhead{\tablenotemark{\rm \scriptsize b}$P_{\star}$} & \colhead{\tablenotemark{\rm \scriptsize c}$S_{\rm max, 0.5}$} & \colhead{\tablenotemark{\rm \scriptsize d}$\langle S \rangle _{\dot{M},0.5}$} &  \colhead{\tablenotemark{\rm \scriptsize e}$\dot{M}_{\zeta, 0.5}$} &  \colhead{\tablenotemark{\rm \scriptsize f}$\langle t_{\rm exp,\ min} \rangle _{\dot{M},0.5}$} & \colhead{\tablenotemark{\rm \scriptsize g}$t_{\rm p}$} &  \colhead{\tablenotemark{\rm \scriptsize h}$\tau_{\rm J}$} & \colhead{\tablenotemark{\rm \scriptsize i}$\dot{M}$} & \colhead{\tablenotemark{\rm \scriptsize j}$\dot{E}$} & \colhead{\tablenotemark{\rm \scriptsize k}$S_{\rm max,[0,2,0.5]}$}\\   
    ($10^{15}$\,G) & (ms) & ($\rm k_B/baryon$)  & ($\rm k_B/baryon$) & ($\rm M_{\odot} \ s^{-1}$) & (s) & (ms) &  (s) & ($\rm M_{\odot} \ s^{-1}$) & ($\rm ergs \ s^{-1}$) &($\rm k_B/baryon$) \\
    & & & (min, max) & & & & & & 
    }

    \startdata
    3 & 200 & 176 & (59, 144) & $8.8\times 10^{-6}$ & $4.6\times 10^{-4}$  & 33 & 2.4 & $2.9 \times 10^{-4}$ & $6.4\times 10^{48}$ & 198 \\
    &  20 & 331 & (54, 227) & $1.1\times 10^{-5}$ & $4.6\times 10^{-4}$  & 32 & 6.4 & $3.0 \times 10^{-4}$ & $2.7\times 10^{49}$ & 480 \\
    & \tablenotemark{\rm \scriptsize \textit{tc}} 20 & 355 & (54, 204) & $1.2\times 10^{-5}$ & $4.6\times 10^{-4}$  & 32 & 6.4 & $3.1 \times 10^{-4}$ & $2.7\times 10^{49}$ & 553 \\
    &  \tablenotemark{\rm \scriptsize \textit{lr}} 20 & 290 & (55, 198) & $1.9\times 10^{-5}$ & $7.5\times 10^{-4}$  & 29 & 7.1 & $2.8 \times 10^{-4}$ & $2.6\times 10^{49}$ & 429 \\
    &  10 & 591 & (51, 390) & $1.3\times 10^{-5}$ & $3.6\times 10^{-4}$  & 26 & 9.6 & $3.3 \times 10^{-4}$ & $6.4\times 10^{49}$ & 1038 \\
    &  5\tablenotemark{\rm \scriptsize \tiny{\#}} & 1077 & (71, 708) & $2.6\times 10^{-5}$ & $1.8\times 10^{-4}$  & 20 & 14.4 & $3.9 \times 10^{-4}$ & $1.7\times 10^{50}$ & 1712 \\ 
    &  \tablenotemark{\rm \scriptsize l}20 & 261 & (55, 168) & $1.3\times 10^{-5}$ & $4.2\times 10^{-4}$  & 28 & 5.5 & $3.9 \times 10^{-4}$ & $3.3\times 10^{49}$ & 414\\
    &  \tablenotemark{\rm \scriptsize l,$hr$}20 & 263 & (54, 187) & $1.1\times 10^{-5}$ & $5.0\times 10^{-4}$  & 28 & 5.8 & $4.0 \times 10^{-4}$ & $3.0\times 10^{49}$ & 383 \\
    &  \tablenotemark{\rm \scriptsize l,n}20 & 299 & (58, 198) & $1.2\times 10^{-5}$ & $4.4\times 10^{-4}$  & 28 & 5.4 & $4.0 \times 10^{-4}$ & $3.5\times 10^{49}$ & 470\\
    &  \tablenotemark{\rm \scriptsize o}200 & 184 & (60, 135) & $9.4\times 10^{-6}$ & $4.7\times 10^{-4}$  & 34 & 2.5 & $2.9 \times 10^{-4}$ & $6.5\times 10^{48}$ & 186\\ &  \tablenotemark{\rm \scriptsize o}20 & 335 & (55, 203) & $1.1\times 10^{-5}$ & $4.8\times 10^{-4}$  & 32 & 6.4 & $3.1 \times 10^{-4}$ & $2.7\times 10^{49}$ & 510\\
    &  \tablenotemark{\rm \scriptsize p}20 & 347 & (55, 201) & $1.1\times 10^{-5}$ & $4.4\times 10^{-4}$  & 32 & 6.3 & $3.1 \times 10^{-4}$ & $2.7\times 10^{49}$ & 501\\
    &  \tablenotemark{\rm \scriptsize n,q}20 & 389 & (56, 246) & $1.0\times 10^{-5}$ & $4.7\times 10^{-4}$  & 39 & 7.7 & $2.4 \times 10^{-4}$ & $2.5\times 10^{49}$ & 772 \\
    &  \tablenotemark{\rm \scriptsize n,q}200 & 189 & (64, 137) & $8.3\times 10^{-6}$ & $4.7\times 10^{-4}$  & 42 & 2.6 & $2.3 \times 10^{-4}$ & $6.1\times 10^{48}$ & 190 \\
    1 & 200 (evol.) & 153 & (52, 132) & \tablenotemark{\rm \scriptsize \tiny{\$}}$1.2\times 10^{-8}$ & $1.4\times 10^{-3}$ & 75 & 9.0 & $2.0\times 10^{-4}$ & $1.2\times 10^{48}$ & 178 \\
     & 20 (evol.) & 229 & (51, 215) & \tablenotemark{\rm \scriptsize \tiny{\$}}$3.6\times 10^{-8}$ & $7.2\times 10^{-4}$ & 69 & 26.0 & $2.1\times 10^{-4}$ & $4.9\times 10^{48}$ & 578 \\
   %0.5  & 20 (evol.) & 207 & (53, 181) & \tablenotemark{\rm \scriptsize \tiny{\$}}$1.7\times 10^{-9}$ & $1.5\times 10^{-3}$ & 210 & 65.0 & $1.4\times 10^{-4}$ & $2.2\times 10^{48}$ & 340 \\
\enddata
\tablenotetext{\rm \scriptsize a}{ Polar magnetic field strength of the PNS.}
\tablenotetext{\rm \scriptsize b}{ Spin period of the PNS.}
\tablenotetext{\rm \scriptsize c}{ Maximum entropy of the outflowing material at the $T=0.5$\,MeV surface over the course of the simulation.}
\tablenotetext{\rm \scriptsize d}{ Average entropy of the outflowing material weighted by $\dot{M}$ within an angle of 3$^{\circ}$ on either side of the equator at the $T=0.5$\,MeV surface. The values in the parentheses show the minimum and maximum of this value over the course of the simulation.}
\tablenotetext{\rm \scriptsize e}{ Time-averaged mass outflow rate of the material with the figure of merit parameter $\zeta \ge\zeta_{\rm crit}$ (equation \ref{zeta_eqn}).} 
\tablenotetext{\rm \scriptsize f}{ Minimum value over the course of the simulation of the average expansion timescale (equation \ref{texp}) of the outflowing material weighted by $\dot{M}$ within an angle of 3$^{\circ}$ on either side of the equator at the $T=0.5$\,MeV surface.}
\tablenotetext{\rm \scriptsize g}{ Time-averaged interval between plasmoids.}
\tablenotetext{\rm \scriptsize h}{ Time-averaged spindown timescale of the PNS (equation \ref{tauj}) measured at a radius $r=50$\,km (see Section \ref{const_lum_sub}).}
\tablenotetext{\rm \scriptsize i}{ Time-averaged mass outflow rate (equation \ref{Mdot}) measured at a radius $r=50$\,km (see Section \ref{const_lum_sub}).}
\tablenotetext{\rm \scriptsize j}{ Time-averaged energy outflow rate (equation \ref{Edot}) measured at a radius $r=1000$\,km (see Section \ref{const_lum_sub}).}
\tablenotetext{\rm \scriptsize k}{ Maximum entropy of the outflowing material in the temperature range $0.2$\,MeV$\le T \le0.5$\,MeV over the course of the simulation.}
\tablenotetext{\rm \scriptsize \textit{tc}}{ Simulation with constant temperature along both radial and $\theta$ directions in the inner boundary ghost zones (see Section \ref{BCs}).}
\tablenotetext{\rm \scriptsize \textit{lr}}{  Simulation with a resolution of ($N_r,\ N_{\theta}$)=($256,\ 128$). All the other simulations are with ($N_r,\ N_{\theta}$)=($512,\ 256$).}
\tablenotetext{\rm \scriptsize l}{ Simulation with second choice inner BCs (see Section \ref{BCs}).}
\tablenotetext{\rm \scriptsize \textit{hr}}{  Simulation with a resolution of ($N_r,\ N_{\theta}$)=($1024,\ 512$). All the other simulations are with ($N_r,\ N_{\theta}$)=($512,\ 256$).}
\tablenotetext{\rm \scriptsize n}{ Simulation with Helmholtz EOS.}
\tablenotetext{\rm \scriptsize o}{ Simulation with a constant density of $10^{12}$\,g cm$^{-3}$ at the inner boundary. All the other inner BCs are as described in Section \ref{BCs}.}
\tablenotetext{\rm \scriptsize p}{ Simulation with radial velocity inner boundary condition set by conserving $\dot{M}$. All the other inner BCs are as described in Section \ref{BCs}.}
\tablenotetext{\rm \scriptsize q}{ Simulation with only charged current neutrino heating/cooling processes and inner boundary conditions set by $\dot{q}=0$ and $Y_{\rm e}$ in ghost zones set equal to the electron fraction in the first active computational zone. All the other inner BCs are as described in Section \ref{BCs}.}
\tablenotetext{\rm \scriptsize \tiny{\#}}{ Simulation in which the Alfv\'en speed and the fast magnetosonic speed exceed the speed of light. Inclusion of relativistic effects is required to confirm this result, which will be our focus in a future work.}
\tablenotetext{\rm \scriptsize \tiny{\$}}{The figure of merit parameter $\zeta$ (equation \ref{zeta_eqn}) does apply to proton-rich winds, but we present this value for comparison (see Section \ref{evol_sub}).}
\end{deluxetable*}

%\vspace{-5mm}
\subsection{Analytic estimates}
\label{scaling_sub}
As described in Section \ref{const_lum_sub}, our simulations show that the PNS rotation rate significantly affects the entropy of the material ejected in plasmoids. We provide analytic estimates of the scaling of entropy in plasmoids as a function of PNS angular frequency of rotation $\Omega_{\star}$. We use force balance at the PNS equator to derive the scaling relations.

The wind pressure is dominated by baryons at the PNS surface, but due to rapid decrease in density outside the PNS surface, the wind pressure is dominated by electrons, positrons, and photons within just a few kilometers from the PNS surface. The wind pressure $P_{\rm w}$ in the regions where plasmoid eruptions occur can be approximated as $P_{\rm w}\approx AT^{4}$, where $T$ is the wind temperature and $A=\frac{11\pi^2}{180}\frac{\rm k_B^4}{\left(\hbar c\right)^3}$ \citep{QW1996}. Plasmoids are ejected when the wind pressure gradient becomes large enough such that the combined force density due to the magnetic field $\bm{F}_{\rm m}$ and the PNS gravity $\bm{F}_{\rm g}$ cannot provide the required centripetal force density $\bm{F}_{\rm c}$ to enable co-rotation of the trapped matter with the PNS. The force density due to the magnetic field is a combination of the magnetic energy density gradient $\bm{F}_{\rm b}\propto \nabla B^2$ and the magnetic tension $|\bm{F}_{\rm t}|\propto \frac{B^2}{R_c}$, where $B$ is the magnitude of the local magnetic field and $R_c$ is the radius of curvature of the magnetic field line. Thus, the force balance condition at the PNS equator just before a plasmoid eruption is,
\begin{equation}
\label{force_balance_eqn}
    \nabla P_{\rm w} + \bm{F}_{\rm c} = \bm{F}_{\rm g} + \bm{F}_{\rm m}.
\end{equation}

Before deriving the scaling relations, it is important to specify the domain of values of $\Omega_{\star}$ where the scalings hold true. Although the force balance condition is true for all values of $\Omega_{\star}$, the derived scaling relations work only if the magnitude of $\bm{F}_{\rm c}$ is comparable to the other forces. We can expect PNS rotation to impact the wind entropy when the PNS is rotating fast enough such that $|\bm{F}_{\rm c}|$ is comparable to the wind pressure gradient at temperatures $T\gtrsim 0.2$\,MeV. In non-magnetized and non-rotating 1D wind models, $T=0.5$\,MeV occurs at about $\sim 25-50$\,km and $T=0.1$\,MeV occurs at $\sim 150-500$\,km from the PNS surface depending on the neutrino luminosity \citep{Thompson2001}. As shown in Figure \ref{entr2d}, rapid PNS rotation causes the $T=0.5$\,MeV surface to come closer to the PNS in the equatorial region. But, as shown in Figure \ref{plasmoids_sequence}, plasmoids also modulate the distance of constant temperature surfaces from the PNS surface. Estimating an average radius of $r\sim 100$\,km for the $T=0.5$\,MeV surface, and a corresponding mass density $\rho \sim 5\times 10^{5}-10^{6}$\,g cm$^{-3}$ (these are typical values in our constant neutrino luminosity simulations) at high neutrino luminosities occurring $\sim 1-2$\,s after the onset of cooling, the condition 
\begin{equation}
\label{omega_crit}
AT^4\sim \rho r^2\Omega_{\star, \rm crit}^2    
\end{equation}
yields $\Omega_{\star, \rm crit}\sim 300-400$\,rad s$^{-1}$ corresponding to $P_{\star, \rm crit}\sim 15-20$\,ms for PNS rotation to affect the wind entropy. 

Although we cannot explore in detail the thermodynamics at $T<0.5$\,MeV due to the approximations to the EOS in this paper (see Section \ref{micro}), slower rotation of $P_{\star}\sim 50$\,ms may be sufficient to significantly affect the thermodynamic conditions in the wind at $T<0.5$\,MeV. As shown in Figure \ref{plasmoids}, the constant temperature surfaces come closer to the PNS as the neutrino luminosities decrease. Depending on the factors by which $\rho$ and $r$ change at a given temperature surface (the factors depend on neutrino luminosities, $B_0$, and PNS spin period), a higher or lower value of $\Omega_{\star}$ is required for PNS rotation to affect the wind entropy as the neutrino luminosities decrease.   

In order to get analytic estimates, we guess that $\bm{F_{\rm m}}$ just before an eruption is a function of only the magnetic field, and that $\bm{F_{\rm m}}$ is independent of $\Omega_{\star}$, which is a reasonable expectation. As an analogy, this is similar to a mass-loaded rotating rubber band. As the rotation rate and/or the mass on the rubber band increase, the tension in the rubber band eventually approaches the breaking point. The breaking point is a function of the material of the rubber band and not the rotation rate.   

From equation \ref{force_balance_eqn}, we thus have the condition that $\nabla P_{\rm w} + \bm{F}_{\rm c} - \bm{F}_{\rm g}$ is independent of $\Omega_{\star}$ just before an eruption at the PNS equator. Considering the radial components at the PNS equator, we get,
\begin{equation}
\label{omc1}
    \frac{d}{d\Omega_{\star}}\left[\frac{\partial P_{\rm w}}{\partial r} + \rho r \Omega_{\star}^2-\frac{GM_{\star}\rho}{r^2}\right] = 0, 
\end{equation}
where $\rho$ is the mass density of the trapped matter, $M_{\star}$ is the PNS mass, $G$ is the gravitational constant, and $r$ is the radius at which the eruption occurs. We have $\frac{\partial P_{\rm w}}{\partial r}=4AT^3\frac{\partial T}{\partial r}$. To satisfy equation \ref{omc1}, we require (under certain assumptions) that each term in the brackets on the left hand side be independent of $\Omega_{\star}$. We prove this assertion in Appendix \ref{appendix}. 

Thus, we have the following conditions just before a plasmoid eruption at a given magnetic field $B$:
\begin{align}
    T^3\frac{\partial T}{\partial r} &\propto \Omega_{\star}^0, \\
    \rho r \Omega_{\star}^2 &\propto \Omega_{\star}^0, \ {\rm and} \\
    \frac{\rho}{r^2} &\propto \Omega_{\star}^0.
\end{align}
From the above conditions, we get, $\rho \propto r^2$, $r\propto \Omega_{\star}^{-2/3}$, and $\rho\propto \Omega_{\star}^{-4/3}$. We can expect $T^3\frac{\partial T}{\partial r}$ to scale as $\sim \frac{T^4}{r}$, thus obtaining $T\propto \Omega_{\star}^{-1/6}$. At temperatures $T\gtrsim 0.5$\,MeV, the entropy $S$ is proportional to $T^3/\rho$ \citep{QW1996}. We thus obtain the following scaling for entropy as a function of $\Omega_{\star}$:
\begin{equation}
\label{entr_scale_eqn}
    S\propto \Omega_{\star}^{5/6}.
\end{equation}

These estimates suggest that the enhancement of entropy in plasmoids with faster PNS rotation is primarily due to lower mass density on the magnetic field lines (we emphasize that the magnetic field lines are not real entities, but are useful in providing a physical intuition). The reason for lower mass density associated with faster PNS rotation can be understood by invoking the mass-loaded rubber band analogy mentioned above. Adding mass to a rotating rubber band stretches it. As mentioned earlier, the maximum tension a rubber band can tolerate depends on the material and not on the total mass added or the the rotation rate. The tension in the rubber band is proportional to both the mass it carries and the rotation rate. As the rotation rate increases, the critical tension is reached for a smaller value of the total mass on the rubber band. This analogy is helpful in getting a physical intuition in the case of plasmoid eruptions. As the PNS rotation rate increases, the total mass trapped on the magnetic field lines at the point of eruption decreases, leading to a smaller mass density on the field lines.

\begin{figure}
\centering
\vspace{3mm}
\includegraphics[width=\linewidth]{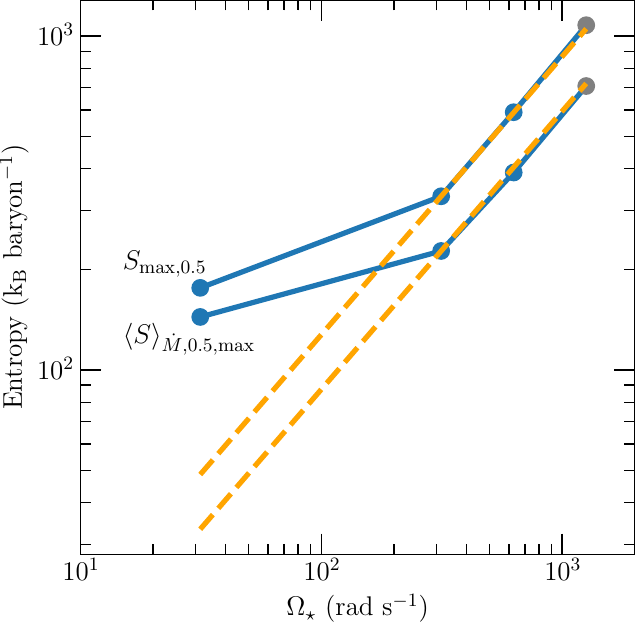}
\caption{The solid blue lines show maximum entropy $S_{\rm max}$ and the maximum of the average entropy weighted by $\dot{M}$ in the zones within $3^{\circ}$ on either side of the PNS equator at the $T=0.5$\,MeV surface as a function of PNS angular frequency $\Omega_{\star}$ at a polar magnetic field strength $B_0=3\times 10^{15}$\,G. The filled circles are the data points from constant luminosity simulations (see Section \ref{const_lum_sub} and Table \ref{table1}). The data point from $P_{\star}=5$\,ms ($\Omega_{\star}=1256$\,rad s$^{-1}$) is shown as a gray colored filled circle to emphasize that the Alfv\'en speed and the fast magnetosonic speed exceed the speed of light in this simulation (refer to the caption of Figure \ref{mdot_zeta}). The dashed orange lines show the $\Omega_{\star}^{5/6}$ scaling of entropy (see the proportionality in \ref{entr_scale_eqn}). We note that this is an approximate verification of the scaling (see Section \ref{scaling_sub} for details).} 
\label{entr_scaling}
\end{figure}

To verify the scaling relations derived above, the region where a field line erupts needs to be identified. Due to the lack of a straight-forward computational condition to identify the point of eruption from the simulation data, we present approximate verification of the entropy scaling, which is sufficient for our purposes in this paper. We verify the scaling at a fixed $T=0.5$\,MeV temperature surface. This approximate verification assumes that the entropy of the material ejected in the plasmoids does not change significantly between the region of eruption and the $T=0.5$\,MeV surface. This is a reasonable approximation because the radial velocity of the material ejected in the plasmoids is $\sim 10^{10}$\,cm s$^{-1}$ as shown in Figures \ref{plasmoids} and \ref{plasmoids_sequence}. Since the material has to travel a distance of at most $\sim 50-100$\,km to reach $T=0.5$\,MeV (because the $T=0.5$\,MeV surface is itself at a distance of $\sim 100$\,km in our constant luminosity simulations), it takes a time of $\lesssim 1$\,ms, which is much smaller than the timescale for plasmoid ejection (see Table \ref{table1}). We can thus safely assume that the entropy enhancement within the $T=0.5$\,MeV surface occurs mostly when the matter is trapped.   

Figure \ref{entr_scaling} shows the maximum entropy $S_{\rm max}$ and the maximum of the average entropy weighted by $\dot{M}$ in the zones within $3^{\circ}$ on either side of the PNS equator at the $T=0.5$\,MeV surface as a function of PNS angular frequency $\Omega_{\star}$ at polar magnetic field strength $B_0=3\times 10^{15}$\,G. Figure \ref{entr_scaling} also shows the $S\propto \Omega_{\star}^{5/6}$ scaling relation derived above. We find excellent agreement between the derived scaling relation and the simulation results for PNS spin periods $P_{\star}\lesssim 20$\,ms as estimated in equation \ref{omega_crit}. We emphasize that the scaling relations fail for $P_{\star}\gtrsim 20$\,ms in the constant luminosity simulations presented in this paper because the PNS is not rotating fast enough at these spin periods for the centripetal force to be comparable to the other forces in the equatorial region.

\subsection{Evolutionary models}
\label{evol_sub}

\begin{figure}
\centering
\vspace{3mm}
\includegraphics[width=\linewidth]{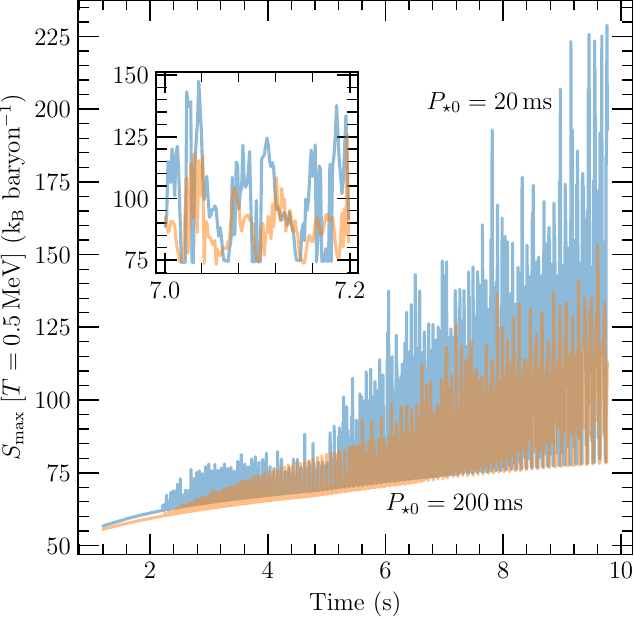}
\caption{Maximum entropy at the $T=0.5$\,MeV surface for the evolutionary models at polar magnetic field strength $B_0=10^{15}$\,G. The blue and orange lines show the profiles for models starting from initial PNS spin periods $P_{\star 0}=20$\,ms and 200\,ms respectively. The inset compares the profiles on a smaller timescale to show differences in the nature of the plasmoids in the two simulations.} 
\label{entr_evol}
\end{figure}

\begin{figure*}
\centering
\vspace{3mm}
\includegraphics[width=\textwidth]{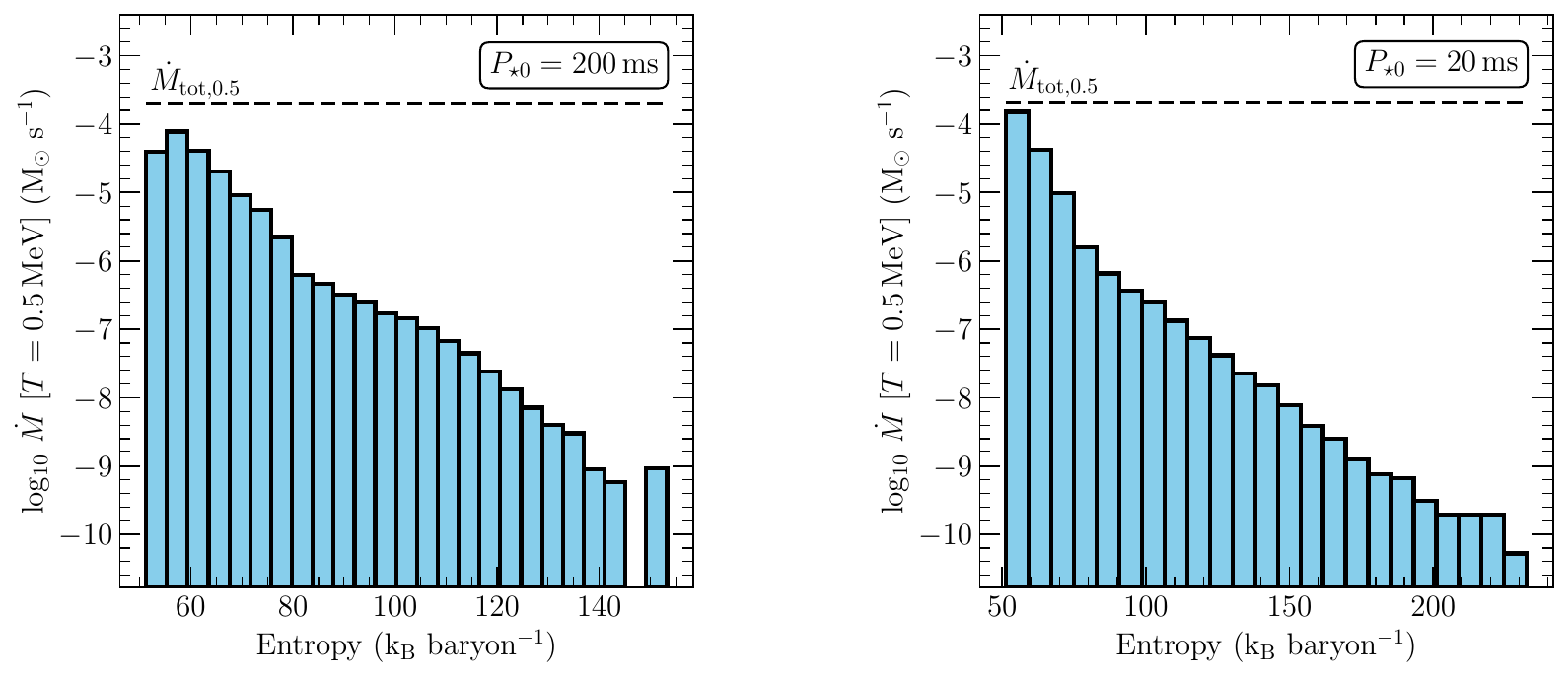}
\caption{Time-averaged mass outflow rate $\dot{M}$ through the $T=0.5$\,MeV surface as a function of entropy for the evolutionary models (see Section \ref{evol_sub}) starting from initial PNS spin periods $P_{\star 0 }=200$\,ms and 20\,ms at a fixed polar magnetic field strength $B_0=10^{15}$\,G, where the time-average has been taken over the entire course of the evolution (note the different entropy limits on the x-axis in the two panels). The dotted black horizontal line shows the total time-averaged $\dot{M}$ through the $T=0.5$\,MeV surface. The values of $\dot{M}_{\rm tot,0.5}$ in this figure are slightly different from the $\dot{M}$ values computed over a spherical surface reported in Table \ref{table1} because the $T=0.5$\,MeV surface is not spherical.}
\label{entr_hist_evol}
\end{figure*}

\begin{figure*}
\centering
\vspace{3mm}
\includegraphics[width=\textwidth]{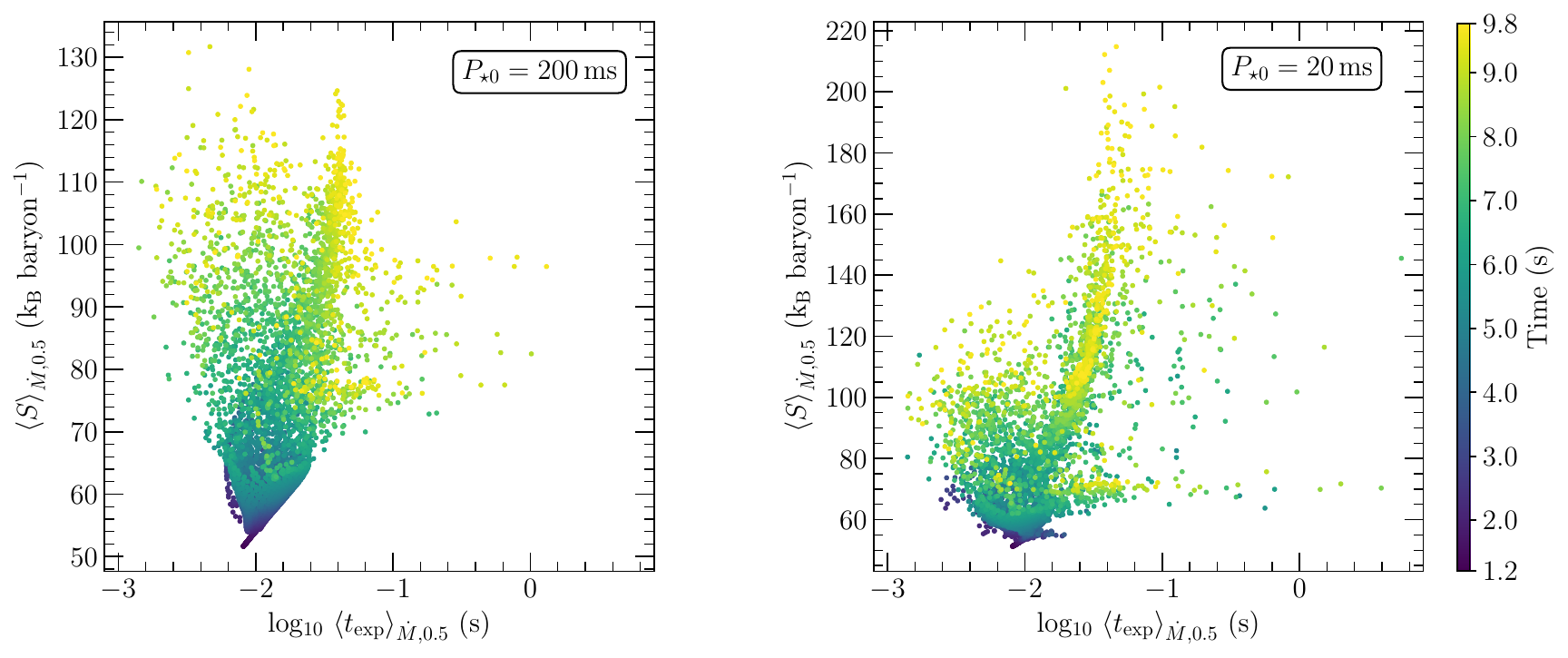}
\caption{Scatter plot of $\dot{M}$ weighted average of entropy and expansion timescale in the zones within 3$^{\circ}$ on either side of the PNS equator at $T=0.5$\,MeV for the evolutionary models starting from initial spin periods $P_{\star 0}=200$\,ms and 20\,ms at $B_0=10^{15}$\,G. Each dot in the scatter plot has been color-coded as a function of time from the onset of the cooling phase and represents the weighted average at one time instant. All the dots together in each scatter plot span a duration of 8.6\,s during the PNS cooling phase. It is evident from these plots that the entropy in plasmoids increases as the PNS cools and the neutrino luminosity decreases at a fixed value of polar magnetic field strength (note the different limits on the y-axis in the two panels).} 
\label{scatter_evol}
\end{figure*}

\begin{figure}
\centering
\vspace{3mm}
\includegraphics[width=\linewidth]{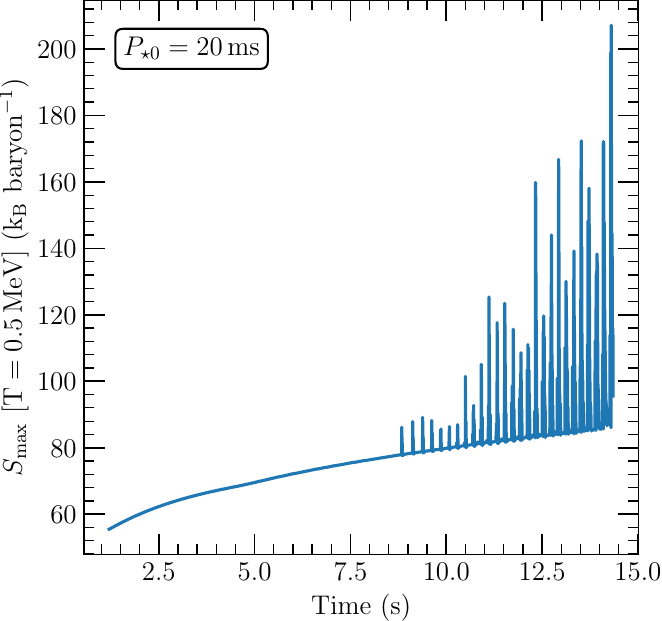}
\caption{Maximum entropy at the $T=0.5$\,MeV surface for the evolutionary model at polar magnetic field strength $B_0=5\times10^{14}$\,G and initial PNS spin period $P_{\star 0}=20$\,ms.} 
\label{entr_evol_5e14}
\end{figure}

\begin{figure*}
\centering
%\vspace{3mm}
\includegraphics[width=\textwidth]{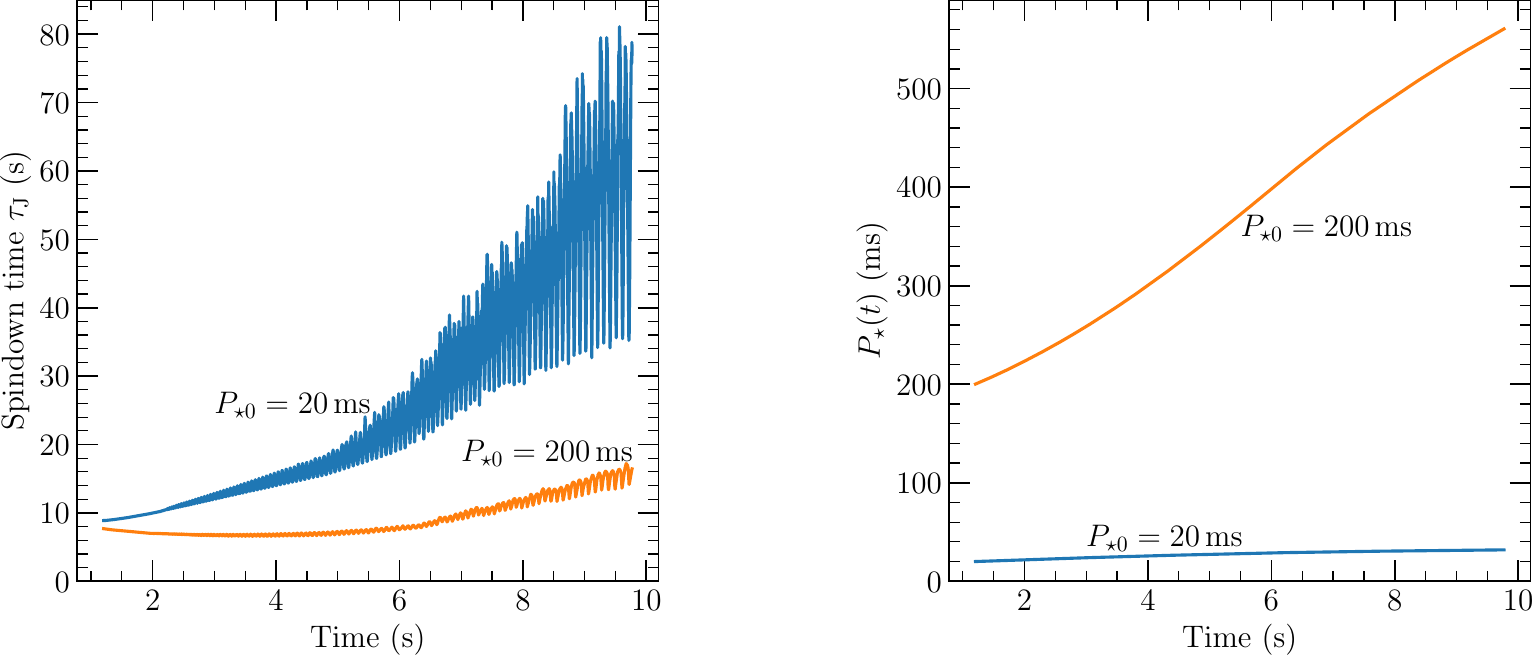}
\caption{The left panel shows the spindown time $\tau_{\rm J}$ and the right panel shows the PNS spin period as a function of time after the onset of the cooling phase for the evolutionary models at polar magnetic field strength $B_0=10^{15}$\,G. The blue and orange lines show the profiles for the models starting from initial PNS spin periods $P_{\star 0}=20$\,ms and 200\,ms respectively.} 
\label{tauj_evol}
\end{figure*}

We have run models with evolving neutrino luminosity and PNS spin period at fixed polar magnetic field strength $B_0=10^{15}$\,G starting from initial spin periods $P_{\star 0}=200$\,ms and 20\,ms, and at $B_0=5\times 10^{14}$\,G starting from $P_{\star 0}=20$\,ms. We use the \cite{Vartanyan2023} cooling model obtained from a 11.75\,M$_{\odot}$ progenitor. The neutrino luminosities and energies span (start and end values in the brackets) $L_{\rm \bar{\nu}_e}=\left[1.2\times 10^{52},1.6\times 10^{51}\right]$\,ergs s$^{-1}$, $L_{\rm \nu_e}=\left[1.3\times 10^{52},1.7\times 10^{51}\right]$\,ergs s$^{-1}$, $L_{\rm \nu_{\mu}}=L_{\rm \bar{\nu}_{\mu}}=L_{\rm \nu_{\tau}}=L_{\rm \bar{\nu}_{\tau}}=\left[8.7\times 10^{51},1.6\times 10^{51}\right]$\,ergs s$^{-1}$, $\epsilon_{\rm \bar{\nu}_e}=\left[16.3,12.0\right]$\,MeV, $\epsilon_{\rm \nu_e}=\left[13.9,11.1\right]$\,MeV, and $\epsilon_{\rm \nu_{\mu}}=\epsilon_{\rm \bar{\nu}_{\mu}}=\epsilon_{\rm \nu_{\tau}}=\epsilon_{\rm \bar{\nu}_{\tau}}=\left[14.9,11.4\right]$\,MeV for the models at $B_0=10^{15}$\,G. The initial luminosities and energies are the same for the model at $B_0=5\times 10^{14}$\,G, but the final luminosities and energies in the order listed above are $1.1 \times 10^{51}$\,ergs s$^{-1}$, $1.1 \times 10^{51}$\,ergs s$^{-1}$, $1.2 \times 10^{51}$\,ergs s$^{-1}$, 11.0\,MeV, 10.1\,MeV, and 10.6\,MeV. The initial luminosities and energies mentioned here occur 1.2\,s after the onset of cooling in the chosen model. We choose a start time of 1.2\,s because this corresponds to the time when the neutrinosphere radii (computed assuming blackbody emission) for all the neutrino species is comparable to the PNS radius of 12\,km in our simulations. We follow the evolution until 10\,s and 14\,s after the onset of the cooling phase for the models at $B_0=10^{15}$\,G and $B_0=5\times10^{14}$\,G respectively.

The cooling model of \cite{Vartanyan2023} provides the data for a duration of 4.8\,s after the onset of cooling. We use the data to obtain a power law fit to neutrino luminosities and extrapolate the neutrino luminosities assuming a constant power law index at times greater than 4.8\,s. We use the blackbody approximation $\epsilon_{\nu} \propto L_{\nu}^{1/4}$ to obtain neutrino energies for times greater than 4.8\,s. For these combinations of neutrino luminosities and energies, we find that the electron fraction $Y_{\rm e}$ is greater than 0.5 throughout the evolution for all the evolutionary models. We find that the asymptotic $Y_{\rm e}$ is 0.54 at the beginning of the evolution which increases as the neutrino luminosities and energies decrease. As the neutrino luminosities decrease, we find plasmoid eruptions from the PNS magnetosphere. Plasmoids modulate the value of asymptotic $Y_{\rm e}$ in the equatorial region and the modulation percentage increases as neutrino luminosities decrease. Towards the end of the evolution, we find that plasmoids modulate asymptotic $Y_{\rm e}$ by $\sim 3\%$.          

Figure \ref{entr_evol} shows the maximum entropy at the $T=0.5$\,MeV surface as a function of time after the onset of cooling for the evolutionary models starting from initial spin periods of 200\,ms and 20\,ms at $B_0=10^{15}$\,G. There are no oscillations in the entropy profiles until a time of $\sim 2.2$\,s in Figure \ref{entr_evol} due to a stable PNS magnetosphere. As the neutrino luminosities and energies decrease, the magnetic field becomes dominant resulting in plasmoid eruptions, which causes oscillations in the entropy profiles. It is evident from Figure \ref{entr_evol} that faster PNS rotation results in higher entropy in the plasmoids, consistent with the results from constant neutrino luminosity simulations (see Section \ref{const_lum_sub}). Figure \ref{entr_hist_evol} shows the distribution of mass outflow rate through the $T=0.5$\,MeV surface as a function of entropy for both the evolutionary models at $B_0=10^{15}$\,G, which again shows that faster PNS rotation yields higher entropy material. Figure \ref{scatter_evol} shows a scatter plot of $\dot{M}$ weighted average of entropy and expansion timescale in the zones within 3$^{\circ}$ on either side of the PNS equator at $T=0.5$\,MeV. It is evident from the panels in Figure \ref{scatter_evol} that the entropy in plasmoids increases as the PNS cools and the neutrino luminosity decreases at a fixed value of polar magnetic field strength. 

Figure \ref{entr_evol_5e14} shows the maximum entropy at $T=0.5$\,MeV as a function of time after the onset cooling for the evolutionary model at $B_0=5\times 10^{14}$\,G and initial PNS spin period $P_{\star 0}=20$\,ms. We find that plasmoids develop 9\,s after the onset of cooling. We find that the entropy in plasmoids increases as the neutrino luminosities and energies decrease.  

We present the results from the evolutionary models in Table \ref{table1} (refer to Section \ref{const_lum_sub} for the description of the parameters in Table \ref{table1}). As mentioned above, the asymptotic electron fraction $Y_{\rm e}$ is $>0.5$ in all the evolutionary simulations, and hence the figure of merit parameter $\zeta$ (equation \ref{zeta_eqn}) is not applicable. However, for comparison, we present the values of $\dot{M}$ with $\zeta \ge \zeta_{\rm crit}$ for the evolutionary models in Table \ref{table1} assuming that the distribution of $\dot{M}$ as a function of entropy and $\zeta$ would be roughly the same even for asymptotic $Y_{\rm e}<0.5$.

Similar to our conclusions from the constant neutrino luminosity simulations, we find even in the evolutionary models that PNS rotation rate strongly influences the wind properties, which can directly affect heavy element nucleosynthesis. We discuss the implications of these results for nucleosynthesis in Section \ref{conclusion}.

The evolutionary models also allow us to explore the spindown of PNSs. In an earlier work \citep{Prasanna2022}, we explored the spindown of `slowly' rotating magnetars with polar magnetic field strength $B_0\gtrsim 10^{15}$\,G and spin period $\gtrsim 100$\,ms at the onset of the cooling phase for a duration of $\sim 3-5$\,s with a cooling model different from the one used in this paper. Here, we explore the spindown of PNSs until $10-14$\,s after the onset of cooling using the \cite{Vartanyan2023} cooling model. The left panel in Figure \ref{tauj_evol} shows the spindown timescale as a function of time after the onset of cooling for the evolutionary models at $B_0=10^{15}$\,G. The difference in the nature of $\tau_{\rm J}$ between the `slowly' rotating case with $P_{\star 0}=200$\,ms and the more rapidly rotating case with $P_{\star 0}=20$\,ms is striking. Apart from the modulation due to plasmoids, the spindown timescale is affected by two competing factors as the neutrino luminosity decreases. First, the mass outflow rate $\dot{M}$ decreases with neutrino luminosity, which nominally would cause $\tau_{\rm J}$ to increase. However, as the neutrino luminosity decreases, the Alfv\'en radius increases which causes $\tau_{\rm J}$ to decrease (see \citealt{Prasanna2022} for further details). These two features combine to produce the $\tau_{\rm J}$ profiles shown.

We find that the first effect always dominates for the evolutionary model with $P_{\star 0}=20$\,ms and $B_0=10^{15}$\,G, which results in an increasing $\tau_{\rm J}$ profile (apart from the modulation due to plasmoids). For the evolutionary model with $P_{\star 0}=200$\,ms and $B_0=10^{15}$\,G, we find that the second effect dominates during the early phases of the evolution resulting in a decreasing $\tau_{\rm J}$ profile. Between 2.5\,s and 6\,s, we find that the two effects roughly cancel each other resulting in an almost constant $\tau_{\rm J}$ profile. During the later phases of evolution, the first effect dominates, resulting in an increasing $\tau_{\rm J}$ profile. Spindown timescales of just a few seconds imply rapid spindown. 

The right panel in Figure \ref{tauj_evol} shows the evolution of the PNS spin period during the cooling phase for the evolutionary models at $B_0=10^{15}$\,G. We find that the the PNS in the evolutionary calculation with an initial spin period of 200\,ms and $B_0=10^{15}$\,G spins down to a period of 561\,ms at the end of 8.6\,s of evolution. Based on the results in \cite{Prasanna2022}, we can expect more rapid spindown at a higher magnetic field strength and/or slower initial rotation period. On the other hand, we find that the PNS in the evolutionary calculation with $P_{\star 0}=20$\,ms and $B_0=10^{15}$\,G spins down to just 32\,ms at the end of 8.6\,s of evolution. These results are consistent with the results in our earlier work \citep{Prasanna2022}. 

We find much slower spindown for the evolutionary model at $P_{\star 0}=20$\,ms and $B_0=5\times10^{14}$\,G, where the PNS spins down to 26\,ms at the end of 13\,s of evolution. Although spindown is not significant for initial PNS spin periods $\lesssim 20$\,ms, we emphasize that rapid spindown accompanies the PNS cooling phase for magnetars born slowly rotating with initial spin periods $\gtrsim 100$\,ms with magnetar strength magnetic fields $\gtrsim 10^{15}$\,G.   

\section{Discussion and conclusions}
\label{conclusion}
In this paper, we explore for the first time the prospects for heavy element nucleosynthesis in winds from highly magnetized and rotating neutron stars. Earlier works have explored the prospects for nucleosynthesis in magnetar winds without including the effects of rotation \citep{Thompson2018, Desai2023} or include the effects of rotation in a static magnetosphere geometry \citep{Vlasov2014, Vlasov2017}. Similar to the earlier works with a dynamic magnetosphere \citep{Thompson2018, Desai2023}, we find dynamical eruptions of high entropy material from the proto-neutron star (PNS) magnetosphere when the magnetic field is strong enough to dominate the wind pressure (see Section \ref{plasmoid_sub} and Figures \ref{plasmoids}, \ref{plasmoids_sequence}). We find for the first time that PNS rotation strongly influences the entropy of material in the plasmoids. We present a set of 2D magnetohydrodynamic (MHD) simulations at constant neutrino luminosity for PNS spin periods $P_{\star}=200$\,ms, 20\,ms, 10\,ms, and 5\,ms at a polar magnetic field strength $B_0=3\times 10^{15}$\,G (see Section \ref{const_lum_sub}). Figure \ref{entr2d} shows the 2D maps of entropy for various values of PNS spin period at a fixed $B_0$, where it is evident that the entropy in plasmoids increases with faster PNS rotation.

We use the figure of merit parameter $\zeta$ (equation \ref{zeta_eqn}) to estimate the potential for $r-$process using the values of entropy, expansion timescale, and electron fraction. Favorable conditions to produce the third $r-$process peak can be achieved in the regions with $\zeta\ge \zeta_{\rm crit}\simeq 8\times 10^{9} \ \rm (k_B \ baryon^{-1})^3 \ s^{-1}$ \citep{Hoffman1997}. Figure \ref{mdot_zeta} shows the mass outflow rate $\dot{M}$ with $\zeta \ge \zeta_{\rm crit}$ through the $T=0.5$\,MeV temperature surface as a function of time for various values of PNS spin period at a fixed $B_0$. Figures \ref{entr_hist} and \ref{zeta_hist} show the time-averaged distribution of $\dot{M}$ through the $T=0.5$\,MeV surface as a function of entropy and $\zeta$ respectively. Figure \ref{const_lum_scatter} shows the scatter plot of $\dot{M}$ weighted average of entropy and expansion timescale for various values of PNS spin period. From all these figures, it is evident that the PNS rotation rate strongly influences the entropy and other wind properties that can directly affect nucleosynthesis in winds.         

We present an analytic scaling relation between the entropy in plasmoids and the PNS angular frequency of rotation $\Omega_{\star}$ (see Section \ref{scaling_sub}). We find that the entropy $S$ in plasmoids scales as $S\propto \Omega_{\star}^{5/6}$ (see the proportionality in \ref{entr_scale_eqn}). Figure \ref{entr_scaling} shows that the derived scaling relation agrees very well with the simulation data when $\Omega_{\star}$ is sufficiently large (see Section \ref{scaling_sub} for details). 

We also present results from simulations with evolving neutrino luminosity and PNS spin period at $B_0=10^{15}$\,G and $B_0=5\times 10^{14}$\,G (see Section \ref{evol_sub}). We use the cooling model from \cite{Vartanyan2023} and follow the evolution until 10\,s and 14\,s after the onset of cooling at $B_0=10^{15}$\,G and $5\times10^{14}$\,G respectively. Figure \ref{entr_evol} shows the maximum entropy as a function of time after the onset of cooling for the evolutionary models at $B_0=10^{15}$\,G. Figure \ref{entr_hist_evol} shows the distribution of $\dot{M}$ through the $T=0.5$\,MeV surface as a function of entropy at $B_0=10^{15}$\,G. Figure \ref{scatter_evol} shows the scatter plot of $\dot{M}$ weighted average of entropy and expansion timescale for the evolutionary models at $B_0=10^{15}$\,G. Figure \ref{entr_evol_5e14} shows the maximum entropy at $T=0.5$\,MeV as a function of time after the onset of cooling at $B_0=5\times 10^{14}$\,G. Similar to our conclusions from the constant neutrino luminosity simulations, we find even from our evolutionary models that PNS rotation rate affects the critical nucleosynthesis parameters in PNS winds. Table \ref{table1} summarises the results from the constant neutrino luminosity simulations and the evolutionary models. 

\begin{figure}
\centering
\vspace{3mm}
\includegraphics[width=\linewidth]{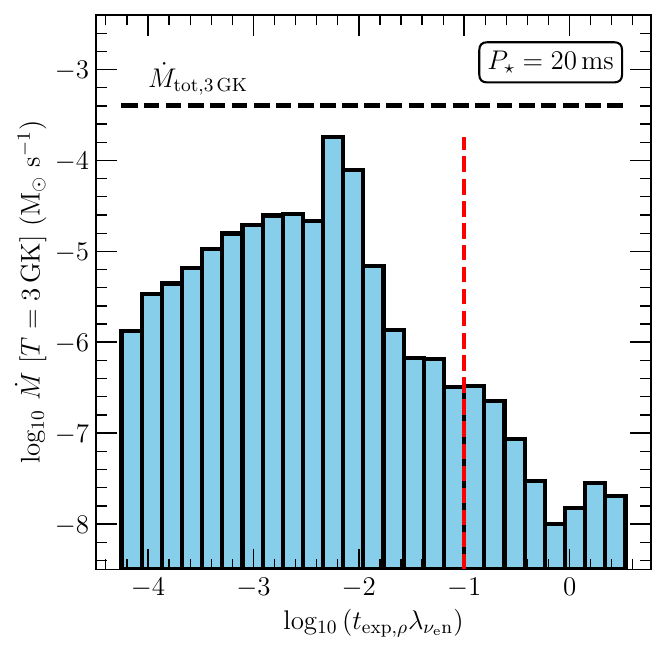}
\caption{Time-averaged mass outflow rate through the $T=3\times 10^{9}$\,K (3\,GK) surface as a function of neutrino exposure defined as the product of expansion timescale $t_{\rm exp, \rho}=\frac{1}{v_r}\left|\frac{1}{\rho}\frac{d\rho}{dr}\right|^{-1}$ (defined using density) and the rate for $\nu_{\rm e}$ absorption on neutrons $\lambda_{\rm \nu_e n}$. The horizontal black dashed line shows the total time-averaged $\dot{M}$ through the $T=3$\,GK surface. The value of $\dot{M}_{\rm tot,3 \ GK}$ in this figure is slightly different from the $\dot{M}$ value computed over a spherical surface reported in Table \ref{table1} because the $T=3$\,GK surface is not spherical. The vertical dashed red line shows the threshold for the $\nu r-$process nucleosynthesis \citep{Xiong2023}. This plot shows data from a constant neutrino luminosity simulation at $B_0=3\times 10^{15}$\,G and PNS spin period $P_{\star}=20$\,ms run with the Helmholtz EOS (the simulation marked with $\rm l,n$ in Table \ref{table1}).} 
\label{exposure}
\end{figure}

Although we find favorable conditions for nucleosynthesis in constant neutrino luminosity simulations and evolutionary simulations, the nucleosynthetic yields depend critically on the electron fraction $Y_{\rm e }$ of the wind. For the neutrino luminosities and assumed energies used in the constant luminosity simulations, we find an asymptotic $Y_{\rm e}\sim 0.45$, which favors the production of $r-$process elements. For PNS spin periods $\gtrsim 10$\,ms in the constant neutrino luminosity simulations, we find that the the time-averaged $\dot{M}$ with $\zeta\ge \zeta_{\rm crit}$ through the $T=0.5$\,MeV surface is $\sim 10^{-5}$\,M$_{\odot}$ s$^{-1}$ (see Table \ref{table1}). Since the timescale for the neutrino luminosities to decrease at the high neutrino luminosities considered in the constant luminosity simulations is $\sim 1-2$\,s, we can expect $\sim 1-2\times 10^{-5}$\,M$_{\odot}$ of high $\zeta$ material to be released during just the first few seconds of the cooling phase at $B_0=3\times 10^{15}$\,G. We can then expect $\sim 2-5\times 10^{-5}$\,M$_{\odot}$ of high $\zeta$ material to be ejected by a magnetar throughout the cooling phase at $B_0=3\times 10^{15}$\,G. Larger masses of high $\zeta$ material may be expected at higher $B_0$. Given a Galactic supernova rate of $\sim 0.02 \ \rm yr^{-1}$ and a fraction $\sim 40-50\%$ of all neutron stars being magnetars \citep{Beniamini2019}, our simulations suggest that we can expect $\sim 2-5\times 10^{-7}$\,M$_{\odot}$ yr$^{-1}$ of high $\zeta$ material to be produced by magnetars. With a conservative estimate accounting for the $\alpha-$rich freezeouts, we can expect $\sim 2-5\times 10^{-8}$\,M$_{\odot}$ yr$^{-1}$ of $r-$process elements with $A>130$ to be produced by magnetars. The production rate of all the heavy $r-$process elements with $A>130$ is $\sim 10^{-7}$\,M$_{\odot}$ yr$^{-1}$ when averaged over the Galaxy's star formation (\citealt{Qian2000}; this reference compares SNe versus neutron star mergers as major $r-$process sources). We reiterate that all the mass estimates with $\zeta\ge \zeta_{\rm crit}$ we have given are at $T=0.5$\,MeV only. We can expect higher yields of high-$\zeta$ material if we consider the wind material at all temperatures. These estimates suggest that magnetar winds are viable candidates for heavy element synthesis.

From the constant neutrino luminosity results presented in Table \ref{table1} and Figure \ref{zeta_hist}, we find that on average, the mass outflow rate of the material with $\zeta \ge \zeta_{\rm crit}$ constitutes $\sim 3-6 \%$ of the total mass outflow rate through the $T=0.5$\,MeV surface. The remaining $\sim 95\%$ of the wind has $\zeta<\zeta_{\rm crit}$. The wind material outside the equatorial region with $\zeta<\zeta_{\rm crit}$ may produce a `weak' $r-$process that can extend to the first and possibly the second $r-$process peak. Detailed simulations with tracer particles are required to study the nucleosynthetic yields that emerge from the distribution of the wind material as a function of $\zeta$.  

The above estimates consider only the $T=0.5$\,MeV surface because the approximations to the EOS (see Section \ref{micro}) do not allow us to explore the wind parameters accurately at $T<0.5$\,MeV. As shown in Table \ref{table1}, we find in the approximate EOS simulations and in simulations with the more accurate Helmholtz EOS that the entropy in the regions with $T<0.5$\,MeV is larger compared to the entropy at $T=0.5$\,MeV. This motivates simulations with the Helmholtz EOS and Lagrangian tracer particles that will enable detailed analysis of the nucleosynthetic yields considering all the regions of the wind. We intend to explore this in detail in a future work.

In our evolutionary models using the \cite{Vartanyan2023} cooling model, we find an asymptotic $Y_{\rm e}>0.5$ throughout the evolution. If the PNS wind is proton-rich, $p-$nuclei can be produced (e.g., \citealt{Pruet2006}). By artificially increasing the entropy by a factor of $2-3$ over their baseline outflow entropy of $55-77$\,$\rm k_B \ baryon^{-1}$, \cite{Pruet2006} find that the nucleosynthesis can extend until $A\approx 170$. They find that even some $r-$process nuclei can be produced in proton-rich winds. These conditions are satisfied in our evolutionary models for a polar magnetic field strength $B_0=10^{15}$\,G, and even a modest $B_0=5\times 10^{14}$\,G, suggesting that PNS winds may be a source of $p-$nuclei.   

Although $Y_{\rm e}>0.5$ in our evolutionary models does not favor $r-$process,  we provide estimates for the $r-$process element yields assuming that the distribution of $\dot{M}$ as a function of entropy and $\zeta$ would be roughly the same even for asymptotic $Y_{\rm e}<0.5$ in these simulations.  Under this assumption, we can estimate using the values in Table \ref{table1} that a magnetar ejects $\sim 1-4\times 10^{-7}$\,M$_{\odot}$ of material with $\zeta \ge \zeta_{\rm crit}$ through the $T=0.5$\,MeV surface in 10\,s during the cooling phase at $B_0=10^{15}$\,G. As mentioned above, we can expect higher values of $\dot{M}$ with $\zeta \ge \zeta_{\rm crit}$ if we consider wind material at all temperatures. For example, we find that $\sim 6\times 10^{-7}$\,M$_{\odot}$ of material satisfies the $\zeta \ge \zeta_{\rm crit}$ condition at $T=0.2$\,MeV in the first 10\,s of evolution at $B_0=10^{15}$\,G starting from an initial spin period of 20\,ms (the estimate at 0.2\,MeV may not be accurate due to the approximations to the EOS). These estimates suggests that magnetars with $B_0\lesssim 10^{15}$\,G cannot account for the entire $r-$process budget of the Galaxy, during the cooling phases explored in these models. As shown in all the evolutionary models, the entropy in plasmoids increases as the neutrino luminosities decrease, and we can expect more favorable nucleosynthesis conditions at lower luminosities. Detailed evaluation throughout the PNS cooling phase with tracer particles is required to assess the nucleosynthetic yields, which will be our focus in a future work.    

\cite{Xiong2023} present a new nucleosynthesis process in which $p-$nuclei and the short-lived nucleus $^{92}\rm Nb$  can be produced in neutron-rich winds via the $\nu r-$process. They show that the $\nu r-$process is enabled if the neutrino exposure, which they define as the product of expansion timescale $t_{\rm exp, \rho}=\frac{1}{v_r}\left|\frac{1}{\rho}\frac{d\rho}{dr}\right|^{-1}$ and the rate for $\nu_{\rm e}$ absorption on neutrons $\lambda_{\rm \nu_e n}$, exceeds 0.1 at the $T\sim3\times 10^{9}$\,K (3\,GK) surface, where they define the expansion time using density. We find that this condition is satisfied in our simulations. At $T\sim 3$\,GK, the approximate analytic EOS \citep{QW1996} is not accurate. Hence, we present results from the simulation run with the Helmholtz EOS. Figure \ref{exposure} shows the time-averaged mass outflow rate through the $T=3$\,GK surface as a function of neutrino exposure (for the simulation marked with $\rm l,n$ in Table \ref{table1}). We find that an average mass outflow rate of $7.5\times 10^{-7}$\,M$_{\odot}$ s$^{-1}$ satisfies the condition $t_{\rm exp,\rho}\lambda_{\rm \nu_e n} \ge 0.1$ at $T=3$\,GK, constituting $\sim 0.2 \%$ of the total mass outflow rate through the 3\,GK surface. This translates to a mass of $\sim 7\times 10^{-7}-1.5\times10^{-6}$\,M$_{\odot}$ with the required conditions for the $\nu r-$process during the first $\sim 1-2$\,s of the PNS cooling phase.

We also study spindown of PNSs using the evolutionary models. Figure \ref{tauj_evol} shows the spindown timescale and the evolution of PNS spin period as a function of time after the onset of cooling at $B_0=10^{15}$\,G. We find that the nature of the profile of spindown time corresponding to the evolutionary model with initial spin period $P_{\star 0}=200$\,ms is significantly different compared to the profile at $P_{\star 0}=20$\,ms (see Section \ref{evol_sub} for details). Similar to the results in our earlier work \citep{Prasanna2022}, we find that `slowly' rotating magnetars born with polar magnetic field strengths $B_0\gtrsim 10^{15}$\,G and initial spin periods $\gtrsim 100$\,ms can spindown significantly during the PNS cooling phase lasting just a few tens of seconds. 

In a work that will follow this paper, we plan to incorporate Lagrangian tracer particles in the simulations to track nucleosynthetic yields. Since our goal in this paper was just to demonstrate that PNS rotation rate significantly affects the nucleosynthesis parameters of the wind, we have chosen to use the approximate analytic form of the general equation of state \citep{QW1996} which is computationally much less expensive than the Helmholtz EOS \citep{Timmes2000}. As mentioned earlier, this approximation to the EOS allows us to explore the wind parameters at only $T\gtrsim 0.5$\,MeV where the approximation is valid. We intend to relax the approximations to the EOS in a future work.

We note that in the constant neutrino luminosity simulation (Section \ref{const_lum_sub}) at a PNS spin period $P_{\star}=5$\,ms and polar magnetic field strength $B_0=3\times10^{15}$\,G, the Alfv\'en speed and the fast magnetosonic speed exceed the speed of light. Since we use non-relativistic physics in our current simulations, inclusion of relativistic effects is required to confirm the result at $P_{\star}=5$\,ms. In the evolutionary models presented in this paper, non-relativistic physics is sufficient as the magnetosonic speeds remain well below the speed of light throughout the evolution. Relativistic simulations are required to study PNS evolution throughout the cooling phase at higher values of $B_0$. This will also be a focus of future work.     

Finally, note that we consider only aligned rotators in this paper. Misaligned rotators where the magnetic axis is tilted with respect to the rotation axis of the magnetar should be investigated. We anticipate several potential effects from misalignment. The first concerns spindown. A limiting case is when the magnetic axis is tilted 90 degrees away from the rotation axis. The polar magnetic field lines are open and the polar field is two times larger than the equatorial magnetic field for a dipole, which suggests that the spindown torque caused by the wind (studied in our earlier papers \citealt{Prasanna2022, Prasanna2023}) will be larger when the field and the rotation axes are misaligned by $90^{\circ}$. A subset of our authors have recently looked at misaligned cases in 3D using an isothermal equation of state with no heating or cooling \citep{Raives2023}. The second potential effect is on the plasmoid dynamics. For 90$^{\circ}$ misalignment, we expect the rotation rate to have less effect on the timing of the eruptions and the entropy in the plasmoids. Interesting physics can be expected when the angles are not aligned. Three dimensional simulations that relax the axisymmetry condition are required to fully analyze the prospects for spindown and nucleosynthesis in the case of misaligned rotators. This will be a focus in a future work.

\section*{Acknowledgments}
\label{section:acknowledgements}
We thank Brian Metzger, Zewei Xiong, Gabriel Mart\'inez-Pinedo, Oliver Just, and Andre Sieverding for helpful discussions. TAT thanks Asif ud-Doula, Brian Metzger, Phil Chang, Niccol\'o Bucciantini, and Eliot Quataert for discussions and collaboration on this and related topics. TP and TAT are supported in part by NASA grant 80NSSC20K0531. We have run our simulations on the Ohio supercomputer \citep{OhioSupercomputerCenter1987}. Parts of the results in this work make use of the colormaps in the CMasher package \citep{CMasher}.

\bibliography{ref}{}
\bibliographystyle{aasjournal}

\appendix
\section{Proof for the assertion after equation 14}\label{appendix}
In principle, the density $\rho$ and temperature $T$ of the trapped matter, the wind pressure $P_{\rm w}$, and the radius $r$ at which the eruption occurs which are a function of the PNS angular frequency of rotation $\Omega_{\star}$ can be expanded as a power series. That is, each of the three terms in the parentheses on the LHS of equation \ref{omc1} can be expanded as follows:
\begin{align}
\label{sums1}
    \frac{\partial P_{\rm w}}{\partial r} &= \sum_k A_k \Omega_{\star}^k \\
\label{sums2}
    \rho r \Omega_{\star}^2 &= \sum_k B_k \Omega_{\star}^k \\
\label{sums3}
    \frac{GM_{\star}\rho}{r^2} &= \sum_k C_k \Omega_{\star}^k, 
\end{align}
where the summation is over all real numbers and $A_{k}$, $B_{k}$, and $C_{k}$ are constant real numbers. Based on the explanation in Section \ref{scaling_sub} where we argue that the sum of the above three terms has to be independent of $\Omega_{\star}$, for any real constant $X$ that is independent of $\Omega_{\star}$, we have,
\begin{equation}
    \sum_k A_k \Omega_{\star}^k + B_k \Omega_{\star}^k - C_k \Omega_{\star}^k = X.
\end{equation}
The above equation requires $A_0 + B_0 - C_0=X$ and $A_k + B_k - C_k=0$ for all $k\neq 0$ (see below for the proof of this statement in the case of only one term in the power series expansion of each term). Since analytic estimates become intractable with infinitely many terms, we assume that $\rho$, $T$, and $r$ can be approximated with a single power law exponent as a function of $\Omega_{\star}$. We emphasize that this approximation is motivated by physical arguments. Based on the mass-loaded rubber band analogy given in Section \ref{scaling_sub}, we can expect the density $\rho$ of trapped matter in the closed zone just before an eruption to decrease with increasing $\Omega_{\star}$. This means that $\rho$ can be expanded in a power series as $\rho=\sum_k \rho_k \Omega_{\star}^k$, where $k<0$ and $\rho_k$ is a constant real number for all $k$. We note that large negative powers (compared to the leading term) can be ignored for sufficiently large values of $\Omega_{\star}$. The remaining terms in the summation can be approximately replaced with a single term as follows: $\rho=\sum_k \rho_k \Omega_{\star}^k \approx \rho_a \Omega_{\star}^a$, where $a$ and $\rho_a$ are constant real numbers. If $\rho$ can be approximated with a single power law as a function of $\Omega_{\star}$, we can expect the same for $r$ and $T$. Based on these arguments, we can approximately replace the summation in equations \ref{sums1}, \ref{sums2}, and \ref{sums3} with a single power law. Under this approximation, we show below that each of the three terms has to be independent of $\Omega_{\star}$ if the sum is independent of $\Omega_{\star}$.

Suppose we have the sum of the three terms as 
\begin{equation}
\label{con1}
A \Omega_{\star}^\alpha + B \Omega_{\star}^\beta + C \Omega_{\star}^\gamma = X,    
\end{equation}
where $A$, $B$, $C$, and $X$ are constants independent of $\Omega_{\star}$. If $\alpha=\beta=\gamma \neq 0$, then the above equation is satisfied only if $A+B+C=0$ and $X=0$. We do not consider this trivial case because $X$ is related to the magnetic tension force in the case of plasmoid ejections and hence is not equal to zero for non-zero magnetic field (see Section \ref{scaling_sub}). 

Let us suppose that there are solutions to equation \ref{con1} for $\alpha \ne 0$, $\beta \ne 0$, and $\gamma \ne 0$ for all $\Omega_{\star}>0$. Since $X$ is independent of $\Omega_{\star}$, doubling the value of $\Omega_{\star}$ on the LHS does not alter the value of $X$. We thus obtain 
\begin{equation}
2^{\alpha}A \Omega_{\star}^\alpha + 2^{\beta}B \Omega_{\star}^\beta + 2^{\gamma}C \Omega_{\star}^\gamma = A \Omega_{\star}^\alpha + B \Omega_{\star}^\beta + C \Omega_{\star}^\gamma.    
\end{equation}
The LHS and the RHS in the above equation are polynomials in $\Omega_{\star}$, and equality is possible only if the coefficients of the respective powers of $\Omega_{\star}$ in the polynomials are equal. This condition requires $\alpha = \beta = \gamma =0$, which is in contradiction with our initial assumption. We can thus conclude that the sum of the three terms can be independent of $\Omega_{\star}$ only if each term is independent of $\Omega_{\star}$.     
\end{document}